\documentclass[sigconf,authorversion]{acmart}

\usepackage[utf8]{inputenc}
\usepackage{colortbl}
\usepackage{makecell}
\usepackage{subfloat}

\usepackage{cleveref}[2012/02/15]%
\crefformat{footnote}{#2\footnotemark[#1]#3}

\newcommand{\bumper}{\begin{center}
		{\rule{0.90\linewidth}{0.4pt}}
\end{center}}

\newcommand{\censorhidden}[1]{}
\newcommand{\censorchange}[2]{#2}
\newcommand{\censor}[1]{\textit{\textless withheld during blind review \textgreater}}

\renewcommand{\censorhidden}[1]{#1}
\renewcommand{\censorchange}[2]{#1}
\renewcommand{\censor}[1]{#1}

\usepackage[inline]{enumitem}

\usepackage{graphicx}
\usepackage{subcaption}
\usepackage{balance}
\usepackage{listings}
\usepackage{url}

\usepackage{float}
\usepackage{enumitem}
\usepackage{xcolor}

\newcommand*\rot{\rotatebox{90}}

\usepackage[para]{threeparttable}

\usepackage{bbding}
\usepackage{pifont}
\usepackage{wasysym}

\usepackage{wasysym}

\usepackage{booktabs}

\usepackage{eurosym}

\newlist{answerlist}{enumerate*}{1}
\setlist*[answerlist,1]{%
	label=\Circle,
} 
\AtBeginDocument{%
	\providecommand\BibTeX{{%
			\normalfont B\kern-0.5em{\scshape i\kern-0.25em b}\kern-0.8em\TeX}}}

\copyrightyear{2020}
\acmYear{2020}
\setcopyright{acmlicensed}\acmConference[ACSAC 2020]{Annual Computer Security Applications Conference}{December 7--11, 2020}{Austin, USA}
\acmBooktitle{Annual Computer Security Applications Conference (ACSAC 2020), December 7--11, 2020, Austin, USA}
\acmPrice{15.00}
\acmDOI{10.1145/3427228.3427243}
\acmISBN{978-1-4503-8858-0/20/12}

\begin{document}

	\title{More Than Just Good Passwords? A Study on Usability and Security Perceptions of Risk-based Authentication}

	\author{Stephan Wiefling}
	\orcid{0000-0001-7917-6065}
	\affiliation{%
		\institution{H-BRS University of Applied Sciences}
	}
	\affiliation{%
		\institution{Ruhr University Bochum}
	}
	\email{stephan.wiefling@h-brs.de}
	
	\author{Markus Dürmuth}
	\affiliation{%
		\institution{Ruhr University Bochum}
		\city{Bochum}
		\country{Germany}}
	\email{markus.duermuth@rub.de}
	
	\author{Luigi Lo Iacono}
	\orcid{0000-0002-7863-0622}
	\affiliation{%
		\institution{H-BRS University of Applied Sciences}
		\city{Sankt Augustin}
		\country{Germany}}
	\email{luigi.lo\_iacono@h-brs.de}

	\begin{abstract}
Risk-based Authentication (RBA) is an adaptive security measure to strengthen password-based authentication. RBA monitors additional features during login, and when observed feature values differ significantly from previously seen ones, users have to provide additional authentication factors such as a verification code. RBA has the potential to offer more usable authentication, but the usability and the security perceptions of RBA are not studied well.

We present the results of a between-group lab study (n=65) to evaluate usability and security perceptions of two RBA variants, one 2FA variant, and password-only authentication. Our study shows with significant results that RBA is considered to be more usable than the studied 2FA variants, while it is perceived as more secure than password-only authentication in general and comparably secure to 2FA in a variety of application types. We also observed RBA usability problems and provide recommendations for mitigation. Our contribution provides a first deeper understanding of the users' perception of RBA and helps to improve RBA implementations for a broader user acceptance. 	\end{abstract}


	\maketitle

\section{Introduction}\label{introduction}

Weaknesses in password-based authentication have been known for a long time~\cite{morris_password_1979,bonneau_science_2012,wang_targeted_2016,von_zezschwitz_honey_2014,das_tangled_2014,florencio_large-scale_2007,dhamija_why_2006}. 
Over the last few years, large-scale password database leaks~\cite{das_tangled_2014} and intelligent password-guessing attacks~\cite{wang_targeted_2016} even revealed new threats for password-based authentication on the internet. Nevertheless, passwords are still the predominant authentication mechanism deployed by online services today~\cite{bonneau_passwords_2015,quermann_state_2018}.

Website owners must implement additional measures to improve account security and to protect their users. Many online services offer Two-factor Authentication (2FA) as such a measure~\cite{quermann_state_2018}%
. However, 2FA proved to be unpopular among users. Although Google introduced and keeps promoting 2FA since 2011~\cite{shah_advanced_2011}, less than 10\% of all active Google users had 2FA enabled in January 2018~\cite{milka_anatomy_2018}. Potential reasons for the low adoption rates could lie in increased burden introduced by continuous demand for two distinct authentication steps~\cite{krol_they_2015} as well as privacy concerns~\cite{venkatadri_investigating_2018}. Risk-based Authentication (RBA)~\cite{freeman_who_2016} is an approach which improves account security with minimal impact on user interaction. Therefore, RBA has the potential to increase password security without degrading usability.

\subsection{Risk-based Authentication (RBA)}

RBA is typically used in addition to password-based authentication. It protects against a rather strong attacker that either knows the correct login credentials (username and password) or is able to guess the correct credentials with a low number of guesses. Examples include \emph{credential stuffing}~\cite{wang_polymorphism_2014}%
, \emph{phishing}~\cite{dhamija_why_2006}%
, or \textit{online guessing attacks}~\cite{wang_targeted_2016}%
. During password entry, the online service monitors and records additional features that are available in the context. Possible features range from network %
or device information %
to biometrics%
. Based on these features, a risk score is estimated which is typically classified into three risk classes (low, medium, high)~\cite{freeman_who_2016,molloy_risk-based_2012,hurkala_architecture_2014}. Based on the risk score and its classification, the online service can perform several actions. If the risk is considered low (e.g., common device, location, and time), access is granted. On a medium risk (e.g., unknown device at a usual location and time), the service typically requests additional information to confirm the claimed identity (e.g., verification of email address~\cite{iaroshevych_improving_2017,shepard_using_2014,freeman_who_2016}). If the risk score is considered high (e.g., unknown device at unrealistic location and uncommon time), the service can block access. This should be a rare event, however, since it will not allow users who are mistakenly classified as a high risk to access their account.

Varying both the exact computation of the risk score and the thresholds separating low, medium, and high risk gives a whole spectrum of variants of RBA. At one end of the spectrum, for an extremely strict risk estimation, re-authentication is requested for \emph{every} login attempt, thus the system appears to users just as 2FA. At the other end of the spectrum, for a very insensitive risk engine, re-authentication is \emph{never} requested, thus the system appears just as password-only authentication. Sensible implementations of RBA are located between those extremes, and require re-authentication only for a %
fraction of the login attempts.  In our study, we will compare two variants of RBA with these alternatives in order to be able to compare across this spectrum.

RBA should not be confused with Implicit Authentication (IA) \cite{khan_usability_2015}, which describes password-less continuous authentication on mobile devices via behavioral biometrics.

The adoption of RBA is still rather limited to few mostly large online services~\cite{milka_anatomy_2018,iaroshevych_improving_2017,allen_risk_2015\censorhidden{,wiefling_is_2019}}. Only five popular online services used RBA in spring 2018~\cite{wiefling_is_2019}. The recommendation by the NIST digital identity guidelines~\cite{grassi_digital_2017} and related research are expected to contribute to a broader usage.

\subsection{Research Questions} \label{section:researchquestions}

The adoption of new approaches and technologies depends on many factors. Among these factors are the usability and security perceptions~\cite{ayaz_building_2019}. Despite its potential and increasing importance, the usability and security perceptions of RBA were not evaluated in literature to date. %
To learn more about these perceptions, we formulated the following research questions. These questions can help to provide answers on how RBA is perceived compared to password-only authentication and equivalent 2FA variants, and if it has the potential to compensate the low adoption rates of 2FA.

\subsubsection*{Usability perceptions:}

\newlist{ULIST}{enumerate}{1}
\setlist[ULIST]{label=\bgroup\bfseries U\arabic*:\egroup, leftmargin=2em, parsep=0em}

\newlist{U2LIST}{enumerate}{2}
\setlist[U2LIST]{label=\bgroup\bfseries \alph*)\egroup,leftmargin=1.3em, parsep=0em}
\begin{ULIST}
	\item \begin{U2LIST}
	    \item How does the usage of RBA affect the user acceptance compared to 2FA? 

	    \item How does the frequency of asking for re-authentication affect the user acceptance of RBA?
    \end{U2LIST}

	\item \begin{U2LIST}
		\item How does the usage of RBA affect the usability regarding the System Usability Scale (SUS)~\cite{brooke_sus:_1996} compared to 2FA?

		\item How does the usability of RBA compare to password-only authentication regarding the SUS?
		\end{U2LIST}
	    
	\item In which context (data to provide, type of website) do users accept RBA?
	
	\item Do users understand why they occasionally have to re-authen\-ticate with RBA?
\end{ULIST}

\subsubsection*{Security perceptions:}

\newlist{SLIST}{enumerate}{1}
\setlist[SLIST]{label=\bgroup\bfseries S\arabic*:\egroup,leftmargin=1.75em, parsep=0em}

\newlist{S2LIST}{enumerate}{2}
\setlist[S2LIST]{label=\bgroup\bfseries \alph*)\egroup,leftmargin=1.55em, parsep=0em}
\begin{SLIST}
	\item \begin{S2LIST}
	    \item How does the security perception of RBA compare to the security perception of 2FA?

		\item How does the usage of RBA affect the security perception compared to password-only authentication?
	\end{S2LIST}
	
	\item \begin{S2LIST}
	   	\item How does the perceived level of protection of RBA compare to the perceived level of protection of 2FA?

		\item How does the usage of RBA affect the perceived level of protection compared to the perceived level of protection of password-only authentication?
	\end{S2LIST}

	\item In which contexts do users feel protected with RBA?
\end{SLIST}

\subsection{Contributions}\label{contributions}

We designed and conducted a between-group lab study with 65 participants and four conditions to evaluate usability and security perceptions of RBA. %
We compared our results with password-only authentication and a 2FA variant. In general, RBA was perceived significantly more secure than password-only authentication. We identified use cases where users significantly preferred RBA over the studied 2FA variants in terms of usability, while having similar security perceptions for both authentication methods. We also show that the way RBA is implemented has an effect on the user acceptance. Beyond that, we discovered potential usability problems that could have a negative effect on the RBA user experience if not addressed by the RBA implementation appropriately.

Our work supports website owners in deciding which authentication method (2FA, RBA, password-only) fits best to the application scenario of their corresponding website. Moreover, our work helps developers to understand how to strengthen password-based authentication without degrading usability. It also provides indications on how to improve the user experience of existing RBA solutions. Finally, researchers obtain insights on how RBA is perceived by users and how this perception compares to other widespread authentication methods.

\section{Study}\label{section:study_website}

To examine and compare different website authentication methods, we created a lab study based on a specifically developed website%
. The website's functionalities were similar to the ones provided by online storage services like Dropbox, Google Drive, or Nextcloud. %
After registration%
, the participant obtained personal storage on the website. The participant could upload, download, share, and delete files. Also, the participant had the possibility to take pictures via webcam. %
These functionalities enabled us to test a website on which participants share and experience sensitive data.

Before accessing the website, the participants were required to log in. After submitting the login credentials%
, each participant perceived one of these four authentication methods (depending on the assigned condition):

\begin{enumerate}[label=(\roman*),leftmargin=2em]
	\item \textbf{2FA}: The participant was prompted for additional authentication after each successful password entry. More specifically, the participant was requested to enter a security code that was sent to the participant's email address.
	\item \textbf{RBA-DEVICE} (RBA-DEV): The participant was prompted for re-authentication via email, as in the 2FA condition, but only in cases
	where the device used for logging in was never used before by the user.
	\item \textbf{RBA-LOCATION} (RBA-LOC): The participant was prompted for re-authentication via email, as in the 2FA condition, but only in cases where the device's location was never seen before for this user.
	\item \textbf{PASSWORD-ONLY} (PW-ONLY): The participant was not prompted for any additional authentication at all.
\end{enumerate}

We chose these four methods and the re-authentication via email based on \censorchange{}{the findings of Wiefling et al.~\cite{wiefling_is_2019} and }our own observations on the state-of-the-art deployments regarding RBA and other popular authentication methods\censorhidden{~\cite{wiefling_is_2019}}. We assumed that the perception of RBA is dependent on its implementation. Since most of the RBA deployments \censorchange{}{investigated by Wiefling et al.~\cite{wiefling_is_2019} }checked for either the device itself or the location of the device, we decided to test the RBA variations RBA-DEV and RBA-LOC.

\subsection{Design Decisions}\label{design-decisions}

Testing RBA in a usability study is difficult for a number of reasons:
\begin{enumerate*}[label=(\roman*)]
    \item RBA is not standardized at the moment. Thus, currently deployed RBA solutions differ widely in dialog design and implementation~\cite{wiefling_is_2019}.
    \item Fundamental properties of RBA are based on user behavior. Using another device or geolocation will have an effect on the RBA experience.
    \item Also, testing location changes is a challenge for a controlled lab study.
    \item %
    RBA differs from 2FA only in cases where login patterns (e.g., location or device) have been used for the login before.
    To compare RBA with 2FA, participants need to experience a difference between these two authentication methods.
\end{enumerate*}

We addressed all these issues in our study design. We decided to conduct a between-group lab study, including device and location changes, %
to observe user reactions under controlled conditions. We involved %
personal devices to create a realistic study scenario.

\begin{figure}
	\centering
	\begin{subfigure}[b]{0.49\linewidth}
		\includegraphics[width=\textwidth]{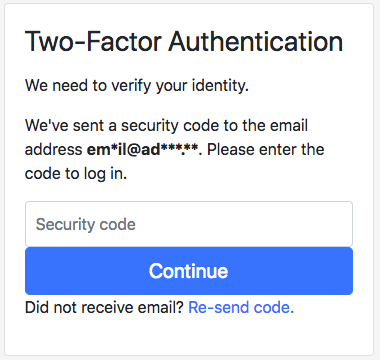}
		\caption{2FA condition}
		\label{fig:auth-2fa}
	\end{subfigure}
	\begin{subfigure}[b]{0.49\linewidth}
		\includegraphics[width=\textwidth]{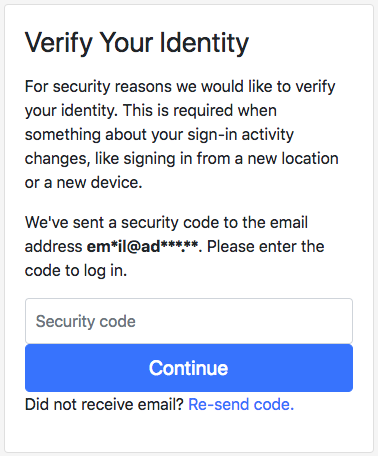}
		\caption{RBA-\{LOC,DEV\} condition}
		\label{fig:auth-rba}
	\end{subfigure}
	\caption{Re-authentication dialogs presented to the %
		study participants for the different login procedures%
		}
	\vspace{-1em}
\end{figure}

We created a generic RBA solution representing state-of-the-art deployments. We decided that our study website requested RBA re-authentication by code-based email address verification since the majority of all online services studied in previous work offered it~\cite{wiefling_is_2019}. %
Online services may offer several 2FA methods in practice, ranging from biometrics to app or code-based solutions. We focused on code-based 2FA via email to ensure comparability with state-of-the-art RBA solutions.
The resulting dialog for RBA-LOC and RBA-DEV (see Figure~\ref{fig:auth-rba}) as well as the sent verification emails were based on the RBA dialogs of Amazon, Facebook, GOG.com, Google, LinkedIn, and Microsoft. %
The 2FA dialog (see Figure \ref{fig:auth-2fa}) is similar to the dialog of LinkedIn%
. We tried to keep the differences between both dialogs at a minimum to mitigate that (completely) different dialog texts could bias the participant's rating in the 2FA and RBA conditions.%

\subsection{Study Design}
\label{section:study}

As the studied authentication methods differ in the login procedure, we required our participants to log in several times. They were asked to solve seven tasks on the study website%
. In these tasks, the participants logged in and out on the website in two different locations using three different devices (2x desktop, 1x mobile device). As a consequence, the participants experienced the corresponding authentication method of the study condition, i.e., participants were asked for re-authentication once (RBA-LOC), twice (RBA-DEV), seven times (2FA), or not at all (PW-ONLY).

We designed the tasks to create an atmosphere where sensitive data is stored and shared on the user account, i.e., confidential company documents and taking a personal picture. Note that pictures are considered more sensitive in Europe compared to other continents~\cite{schomakers_internet_2019}. We assumed that with increased sensitive and personal data, including using a  personal email account and laptop, this would increase the participant's immersion into the study scenario%
. We made this assumption as it is a common observation when studying user authentication that the perceived account value has an influence on user's actions and perceptions.

We introduced two room changes %
during the study to simulate a change of physical location. To strengthen the impression of a location change, both rooms had a very different appearance. \emph{Room A} looked like a typical office room, with white wall and grey furniture colors. \emph{Room B}, our usability lab, looked similar to a living room or hotel room and had warm wall and furniture colors to create a pleasant atmosphere.

\subsection{Study Setup}\label{experiment-procedure}

The between-group lab study consisted of four conditions: 2FA, RBA-DEV, RBA-LOC, and PW-ONLY. All participants were randomly assigned to one of the four conditions while %
balancing genders in each group as far as possible. The study consisted of three stages (task solving, exit survey, and semi-structured interview). The study website was reachable via HTTPS at an internet domain name not connected with our university to mitigate social desirability bias and to increase perceived data sensitivity, i.e., participants don't know where the data is stored.%

The study conductor stayed outside in an observation room next to the study rooms. The conductor could observe the participants' facial reactions as well as display contents of the devices inside \emph{room B} via a streamed video recording.

After solving the tasks, participants answered a survey on a tablet PC. The survey covered five-point Likert scale questions on usability and security perceptions of the login procedure. We integrated several measures into the survey to mitigate known biases and to check the quality of our results (see Section~\ref{exit-survey}%
).

After the survey, we conducted a semi-structured interview with the participants to gain qualitative feedback on both impressions and personal experiences regarding the tested authentication method. We describe the three study stages in detail below.

\subsubsection{Study Procedure}

The study started in \emph{room A} to introduce a typical use case scenario for our participants. The room contained the task sheet, a USB flash drive (containing a presentation and meeting minutes), and a button to call the study conductor in case of questions or support. The study conductors introduced the website as an external cloud storage service. After signing the consent form, the study conductor made it clear to the participants that there were no ``right'' or ``wrong'' answers or actions, and that we test the website not the participants. We did all this to mitigate social desirability bias and to make our participants feel comfortable. The participants were asked to think aloud during the tasks in order to obtain qualitative feedback, especially while they experienced the re-authentication. %
To mitigate fatigue biases, we designed the study to keep the required time for each of the three study stages low (15-20 minutes). All study conductors followed a study script containing all instructions and required materials%
.
All study tasks were printed on sheets of paper, one for each task. The participants turned to the next sheet as soon as each task was completed.

\subsubsection{Study Tasks}\label{task-design}
Participants were asked to bring their private laptop and, if required for accessing personal email, their smartphones to the study. We informed the participants that they were required to use their personal email address for registration on the study website. To avoid bias, we did not mention that this email address was possibly also used for authentication purposes.%

The tasks %
were designed to represent typical situations in working life%
.
Table~\ref{tab:studytasks} gives an overview of the tasks and when re-authenti\-cation was requested in which condition. Note that some real-world online services trigger RBA with slight changes of the IP address, even at the same geolocation~\cite{wiefling_is_2019}. Since all devices involved had individual IP addresses in the study, the tested scenarios are realistic ones.

\renewcommand{\arraystretch}{1.0}
\begin{table}[t]
  \centering
    \caption{Overview of the study tasks and when re-authentication was requested for RBA-LOC, RBA-DEV and 2FA conditions. No re-authentication was requested for the PW-ONLY condition. Room A and the laptop are known to the RBA system as a common context.}
    \begin{threeparttable}
	    \resizebox{\linewidth}{!}{%
	    	\begin{tabular}{@{}rllllll@{}}
	    \toprule[1pt]
	    \# & Task & Room  & Device & \multicolumn{3}{c@{}}{Re-authentication requested} \\
	          & &       &       & RBA-LOC & RBA-DEV & 2FA \\
	    \midrule
	    1 & Register     & A & Laptop & \Circle    & \Circle    & \CIRCLE \\
	    2 & File Upload & A & Laptop & \Circle    & \Circle    & \CIRCLE \\
	    3 & File Download & B & Desktop PC & \CIRCLE   & \CIRCLE   & \CIRCLE \\
	    4 & Open Report & B & Desktop PC & \Circle    & \Circle    & \CIRCLE \\
	    5 & Take Picture & B & Desktop PC & \Circle    & \Circle    & \CIRCLE \\
	    6 & Open File & B & Tablet PC & \Circle    & \CIRCLE   & \CIRCLE \\
	    7 & Delete Data & A & Laptop & \Circle    & \Circle    & \CIRCLE \\
	    \bottomrule[1pt]
	    \end{tabular}%
   	}
		\begin{tablenotes}
		    \small
			\item[\CIRCLE] Requested
			\item[\Circle] Not requested
		\end{tablenotes}
	\end{threeparttable}
  \label{tab:studytasks}
\end{table}

After the study conductor left the room, the participants solved two tasks using their private laptop in \emph{room A}. The tasks introduced the story that the participants are preparing for a meeting at an external business partner. Following that, they registered on the study website (\textbf{task one}), and uploaded a presentation and meeting minutes (\textbf{task two}).

After the two tasks, %
the participants were asked to leave the room, leaving their personal laptop inside the room. \emph{Room A} was locked and the participants were brought into \emph{room B}%
. The workplace was a desk containing a desktop PC (Windows 10 and Chrome browser) with a display, a webcam mounted on it, keyboard and mouse as well as a button to call the study conductor. A tablet PC (Asus Nexus 7, Android 6.0.1 with Chrome browser) was hidden inside the right drawer of the desk. The study conductor left the room and the participants solved three tasks on the desktop PC.

In \textbf{task three}, the participants imagined that they traveled close to the business partner but forgot their laptop at home. Therefore, they entered a (fictional) internet cafe to
download the presentation (uploaded during task two) on a USB flash drive. %
We chose this task to make our participants log in at an unknown device at an unknown location. In \textbf{task four}, the participants
\begin{enumerate*}[label=(\roman*)]
	\item opened a business report (marked as confidential), shared by colleagues on the website,
	\item looked for a quarterly figure in this report and
	\item sent this figure with their personal email client to the email address of a (fictional) business partner.
\end{enumerate*} 
We chose the task to make the participants get in contact with sensitive data%
. In \textbf{task five}, a colleague requested a portrait picture for a company presentation, so the participants took and shared a picture of themselves with this colleague. We chose this task to make our participants share personal data%
.
In \textbf{task six}, the participants
\begin{enumerate*}[label=(\roman*)]
	\item got the tablet PC out of the drawer and
	\item opened the meeting minutes on the tablet PC
\end{enumerate*}
to prepare for the meeting with the business partner.
We chose this task to make our participants log in at an unknown device at the same location.

After the task, %
the participants were brought to \emph{room A} again and solved the final \textbf{task seven} on their laptop. In this task, the participants arrived at home again and deleted the personal data and the user account from the website. We chose this task to make our participants log in at a familiar device at a familiar location (and especially for those in the RBA conditions: to experience that the website recognized them in this common context).

\subsubsection{Exit Survey}\label{exit-survey}

Following the tasks, participants answered a %
survey on a tablet PC to provide quantitative feedback on the authentication methods. The survey consisted of five-point Likert scale questions regarding the user's usability and security perceptions%
. We balanced all survey questions to mitigate social desirability bias~\cite{shaeffer_comparing_2005}. The order of questions and subquestions varied randomly for each participant to randomly distribute ordering effects~\cite{kalton_effect_1982}. Also, the Likert scale direction varied for a randomly selected half of participants in each condition to randomly distribute response order bias~\cite{chan_response-order_1991,hartley_thoughts_2014}.

The first part of the survey consisted of two SUS questionnaires~\cite{brooke_sus:_1996}. We changed the word ``system'' in these questionnaires to ``website'' and ``login procedure'' respectively to explicitly evaluate the usability of the website and the perceived authentication method. We added the SUS questions with the website wording since we briefed our participants that we test a website. The approach to change the SUS wordings is similar to Khan et al.~\cite{khan_usability_2015}. In contrast to them, we left SUS item five\footnote{\textit{``I found the various functions in this system were well integrated''}} inside our questionnaires since we tested a visible user interface%
. We calculated the SUS score as defined in Brooke~\cite{brooke_sus:_1996}. The SUS questionnaire contains attention checks in the form of pairs of related questions with opposite wording %
to verify the quality of our results.

The second part of the survey contained questions on the personal perceptions of the authentication method. Questions ranged from the perceived security and level of protection to the perception and acceptance of the corresponding login method (general and on specific types of websites). Members of the 2FA and RBA conditions also answered questions on understanding, perception, and acceptance of the re-authentication (general and in specific scenarios). We omitted the re-authentication questions for the PW-ONLY condition since the members of this condition were not asked for re-authentication in the study. Some of the survey questions were based on Agarwal et al.~\cite{agarwal_ask_2016} and Khan et al.~\cite{khan_usability_2015}. However, we balanced all of these questions since we found that the original questions could bias participants due to their one-sided, non-neutral wording (e.g., \textit{``How \textbf{annoying} were ...''}~\cite{khan_usability_2015} or \textit{``How \textbf{obstructive} was ...''}~\cite{agarwal_ask_2016}). We chose the uniform wording ``login method'' inside the questions instead of the terms ``scheme''~\cite{agarwal_ask_2016} and ``method''~\cite{khan_usability_2015} since we found this wording easier to understand for our participants.

The survey concluded with basic demographical questions.

\subsubsection{Semi-structured Interview}

Following the exit survey, we conducted a semi-structured interview with the participants%
. We told them that they would not have to answer a question if they did not want to. At the beginning, we asked website-related questions to distract from our actual purpose of the study. Then, we asked questions regarding the login procedure to gain insights on how participants perceived the corresponding authentication method. These questions ranged from likes and dislikes of the login method, their desired changes and security perceptions to suggestions for alternative authentication methods. Members of the 2FA and RBA conditions were additionally asked to explain the login procedure in their own words and to share their personal experiences with similar login procedures. We did this to verify whether they understood why they were asked for re-authentication. Similar to the exit survey, we took some of the questions by Khan et al.~\cite{khan_usability_2015} and Agarwal et al.~\cite{agarwal_ask_2016} into consideration.

\subsection{Data Collection}

To answer our research questions, we collected the following data:
\begin{enumerate*}[label=(\roman*)]
    \item \textbf{Audio and Video:} We recorded a video of the participant's face inside \emph{room B} as well as the screen content of desktop and mobile device (tablet PC). Personal data was automatically censored on the video recording. We also recorded audio of the participant while thinking aloud.
    \item \textbf{Authentication Time:} We recorded the time needed to authenticate on the study website. For this reason, the website stored timestamps of when 
    the first character was entered into the login form and
    the first page was loaded in logged in state.
    We calculated the authentication time as the difference between the two values.
    \item \textbf{Exit Survey:} The survey answers were collected and stored digitally after finishing the survey. %
    \item \textbf{Semi-structured Interview:} We recorded the questions and answers as audio files and transcribed them afterwards.
\end{enumerate*}

\subsection{Ethical Considerations}\label{ethics}

We discovered potential ethical issues while planning the study. Below, we describe these issues and how we addressed them.

\subsubsection{Personal Data on Video}\label{ethics-personal-data}

When requested for re-authentica\-tion, participants had to log into their personal email account to open the re-authentication email. However, there was a risk that contents of other emails were recorded on video when deciding to open this email on the desktop PC. Also, personal email addresses and passwords %
could have been recorded.

To solve this issue, we developed an automatic process to hide personal data from the video recording and stream: The video content was censored automatically (white bar across the entire video) whenever \begin{enumerate*}[label=(\roman*)]
\item a login form was visible or
\item our study website was not focused.
\end{enumerate*}
As a result, login data as well as contents of other browser tabs (e.g.,~the email account) were neither recorded on video nor visible on the video stream. We tested and improved the automatic process over a three week period. We did this to make sure that all device and browser-based use cases are covered, making our process as accurate as possible.

We briefed the participants explicitly about this automatic procedure before the study %
to make them feel comfortable. We also offered the participants to view and inspect the recorded video after the study and to request deletion of the video. One participant made use of that possibility, which underlines that this is an important ethical consideration.

\subsubsection{``Deception''}

We instructed the participants before the study that we evaluate a website. We did not disclose them at the time that we were actually testing authentication methods. However, since the authentication was also part of the website, we considered this deception to be non-critical. We debriefed the participants after the study and revealed them the real purpose of the study.

\subsubsection{Further Precautions}

Besides the automatic process to censor personally identifiable information (PII) on video, we offered our participants additional \textbf{privacy} and \textbf{pseudonymity}, including among others:
\begin{enumerate*}[label=(\roman*)]
	\item \textbf{Login data}: The login credentials, hashed with scrypt~\cite{percival_scrypt_2016}, %
	as well as the picture were only stored during the study and deleted afterwards. %
	\item \textbf{Non-linkability of PII}: After deletion of email address and password, the participants could only be identified by a random sequence of characters and numbers (token).
	\item \textbf{Storage}: All study data was stored on encrypted external mobile hard drives. Only the study conductors had the decryption password, i.e., access to this hard drive.
\end{enumerate*}

The participants were informed by all these procedures and signed a consent form (\textbf{informed consent}). Participants were informed that they could withdraw the study anytime. All survey questions offered a ``don't know'' option.

We did not have a formal IRB process at TH Köln, where we conducted this study. But besides our ethical considerations above, we made sure to minimize potential harm by complying with the ethics code of \censorchange{the German Sociological Association (DGS)}{a nationwide sociological association} as well as the standards of good scientific practice of \censorchange{the German Research Foundation (DFG)}{a nationwide research funding organisation\footnote{\label{footnote:orgname-omitted}Organization names omitted during blind review}}. We also made sure to comply with the terms of the EU General Data Protection Regulation.

\subsection{Piloting}\label{pilot-study}

We piloted the study with three participants to verify and optimize our study procedure. %
In contrast to the final study, we asked the participants to think aloud while answering the exit survey. This helped us to understand how participants interpret the context of the survey questions.
Minor adjustments to survey question wordings were done as a result of piloting.

\subsection{Recruiting}

Our study required participants using online services with private data. Knowledge in neither 2FA nor RBA was not required. We recruited participants via emails sent to mailing lists of faculties in social sciences, biology, medicine, and humanities of \censorchange{University of Cologne}{University A}, and architecture, communication sciences, and engineering faculties of \censorchange{TH Köln}{University B\cref{footnote:orgname-omitted}}. We also put up posters in the corresponding university faculties and advertised on a local radio station targeting a young audience to recruit for the study. We did this to investigate a broad sample of digital natives%
. We took care and selected only participants %
that did not attend any information security lectures to mitigate bias. We mentioned in the recruiting email and poster that the study is about testing a website and that the study lasts about one hour (i.e., $3 \cdot 20$ minutes). %
Among all participants we drew six Amazon gift cards worth 25\euro ~each. We also offered candy bars and drinks for the participants' personal well-being during the study.

\section{Results} \label{section:results}

The study took place between December 2018 and February 2020 and was completed with 65 participants (17 in the PW-ONLY condition, 16 each in the three other conditions)%
.
68 participated but three of them experienced problems with the website or forgot to log out between tasks, which is why they were excluded from the results. The participants were between 19 and 33 years old (mean: 24.57, SD: 3.22). 17 participants were female, 47 were male, and one chose not to state the gender%
. %
RBA-DEV had five female participants, all remaining conditions had four female participants. All study sessions lasted 50 minutes at a maximum.

For the survey data, we used Kruskal-Wallis (K-W) tests for the omnibus cases and Dunn's multiple comparison test with Bonferroni correction (Dunn-Bonferroni) for post-hoc analysis. For the timing comparison (with and without re-authentication), we used Mann-Whitney-U (MWU) tests to compare the statistical difference between the two conditions. We set 0.05 as our threshold for statistical significance (i.e., p$<$0.05 is significant).%

For the semi-structured interview, we pattern-coded the respon\-ses using inductive coding: The answers were read and observed patterns were added to the codebook. After that, the answers were coded into the patterns independently by two researchers of our research group. If both researchers coded an answer differently, a third researcher did the final decision. For the coding%
, we achieved Cohen's Kappa $\kappa$ = 0.82, which is within the acceptable range of coding agreement~\cite{mchugh_interrater_2012}.

Below, we present the qualitative and quantitative study results ordered by our research questions. A discussion follows after presenting the results of each research question.

\subsection{Usability Perceptions}

In this section, we compare the usability of the studied RBA, 2FA, and password-only authentication schemes. Besides the general user acceptance, we identify contexts in which users prefer RBA to 2FA and investigate whether users understand RBA's re-authentica\-tion requests.

\subsubsection{User Acceptance and SUS (U1, U2)} \label{subsection:user_acceptance}

\begin{figure}[t]
	\centering
	\includegraphics[width=0.99\linewidth]{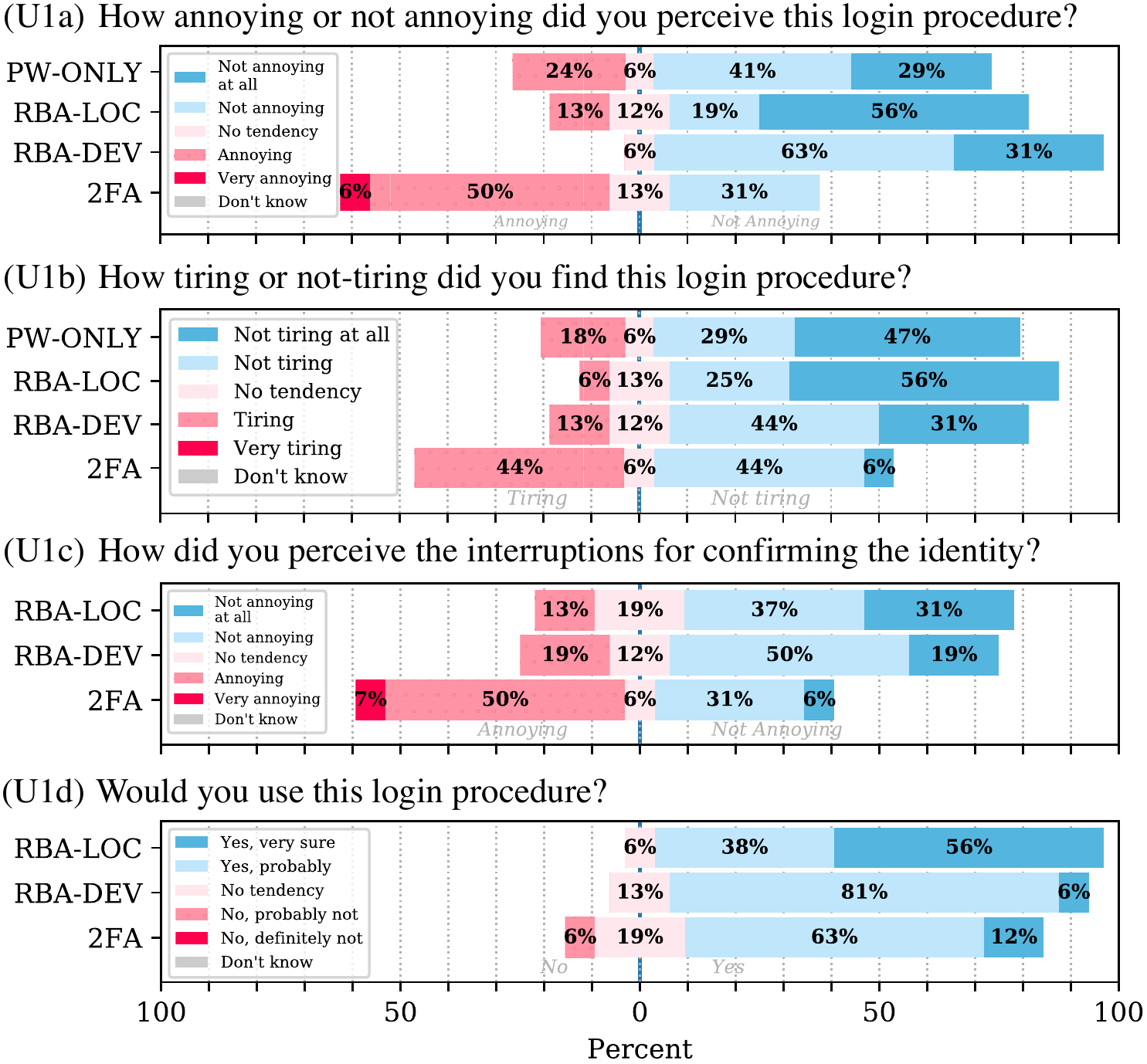}
	\vspace{-1em}
	\caption{%
	Responses to the user acceptance questions (U1)%
	}
	\label{fig:plots_user_acceptance}
\end{figure}

In the exit survey, the participants responded to several questions regarding the acceptance of the corresponding login method (see Figure~\ref{fig:plots_user_acceptance}). There were no significant differences between PW-ONLY and the other three conditions.
However, the participants perceived RBA significantly less annoying than 2FA (RBA-LOC/2FA: p=0.001; RBA-DEV/2FA: p=0.0022). The participants also found RBA-LOC significantly less tiring than 2FA (p=0.0122) and its interruptions significantly less annoying than those of 2FA (p=0.0331). %
The majority of the RBA and 2FA participants agreed with the question of whether they would use their login procedure. RBA-LOC group members, who had to do less re-authentication than those of RBA-DEV and 2FA, gave significantly higher ratings than those of RBA-DEV and 2FA regarding that question (RBA-DEV/RBA-LOC: p=0.026; RBA-DEV/2FA: p=0.0117).

The participants also answered two adjusted SUS surveys. The surveys contained questions about the authentication method %
and the study website respectively. 

	\begin{figure}
		\centering
		\includegraphics[width=0.85\linewidth]{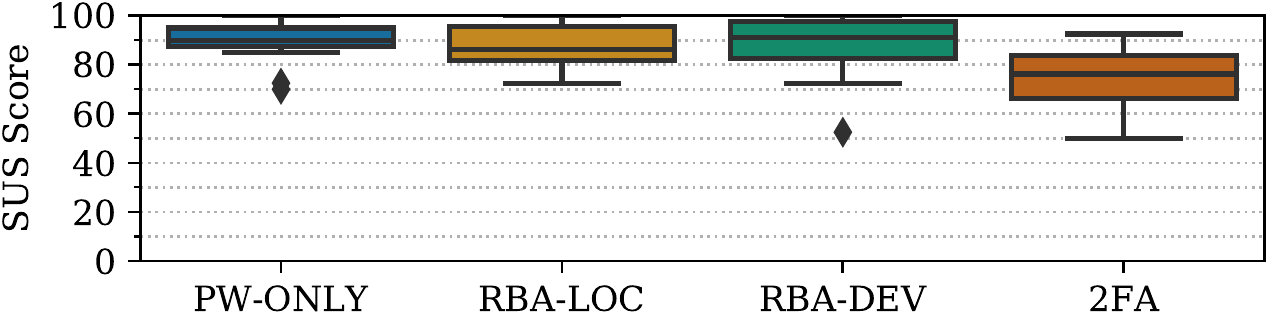}
		\caption{Login procedure usability (U2): Box plot showing the SUS score results for the study conditions. PW-ONLY and RBA-DEV received significantly higher scores than 2FA.%
		}
		\label{fig:resultssuslogin}
	\end{figure}

With a median SUS score above 80, the PW-ONLY and RBA authentication methods can be considered grade A usability~\cite{sauro_quantifying_2012} (see Figure~\ref{fig:resultssuslogin}). With a median SUS score of 76.25, 2FA can be considered grade B usability. The SUS scores of PW-ONLY and RBA-DEV are significantly higher than those of 2FA (see Table~\ref{tab:pvalues-sus}). %
PW-ONLY, RBA-LOC, and RBA-DEV also received significantly more positive ratings than 2FA in some of the SUS questions%
: Participants rated 2FA significantly more cumbersome to use and significantly more unnecessarily complex compared to PW-ONLY and both RBA conditions. Participants would use both RBA variations significantly more frequently than 2FA. PW-ONLY was rated significantly easier to use than 2FA.

\renewcommand{\arraystretch}{1.3}
\renewcommand{\arraystretch}{1.0}
\begin{table}[t]
  \caption{Significant (bold) K-W and Dunn-Bonferroni p-values for the SUS score and SUS questions. We excluded p-values greater than 0.2.}
  \centering
  \resizebox{\linewidth}{!}{
    \begin{tabular}{@{}l|r|rrrr@{}}
      & 	& 2FA/ & 2FA/ & 2FA/ & RBA-L/ \\
      & K-W & PW & RBA-L & RBA-D & RBA-D \\
    \midrule
    SUS score & \textbf{0.0030} & \textbf{0.0093} & 0.0523 & \textbf{0.0073} & - \\
    \midrule
    Use more frequently & \textbf{0.0041} & - & \textbf{0.0185} & \textbf{0.0078} & - \\
    Unnecessarily complex & \textbf{0.0003} & \textbf{0.0005} & \textbf{0.0420} & \textbf{0.0026} & - \\
    Easy to use & \textbf{0.0054} & \textbf{0.0034} & 0.1084 & 0.0964 & - \\
    Cumbersome to use & \textbf{0.0002} & \textbf{0.0005} & \textbf{0.0049} & \textbf{0.0027} & - \\
    \end{tabular}%
	}
  \label{tab:pvalues-sus}%
\end{table}%

Concluding the results, the user acceptance of RBA is in some cases significantly higher than 2FA. For the remaining cases, the user acceptance of RBA is not significantly lower than 2FA.
In addition, RBA-DEV is perceived significantly more usable than 2FA regarding the SUS score. RBA-LOC and RBA-DEV are perceived significantly more usable than 2FA regarding the answers of the SUS questions.
As the main difference of the studied schemes is the amount and frequency of required authentication, we conclude that less requests for re-authentication are accepted significantly higher than more of them%
. Since PW-ONLY also received a significantly more positive rating than 2FA, RBA is comparable to password-only authentication regarding the SUS score and parts of the SUS question answers.

\paragraph{Discussion:}
RBA participants were asked less often for re-authen\-ti\-cation than those of 2FA. We conclude that this was the main reason why RBA and PW-ONLY outweighed 2FA in terms of usability and user acceptance, as 2FA participants mentioned this as well:

\begin{quote}
	\textit{``It was very cumbersome to log in to the email account every time. Especially, if you are not at your own computer, but somewhere else.%
	''} (P15)
\end{quote}

When asked for re-authentication, participants needed significantly more time to authenticate than without re-authentication, due to the requested additional step (MWU: U=1358.5; p$\ll$0.0001, \emph{without}: mean=8~s; median=10.98~s; SD=8.49~s, \emph{with}: mean=59.22~s; median=42~s; SD=55.1~s). Therefore, frequent logins increased the total authentication time and decreased usability and user acceptance%
. One participant of RBA-DEV mentioned this in the semi-structured interview:
 
 \begin{quote}
 	 \textit{``[I liked that] when I was using the same device that I didn't have to authenticate twice by email.''} (P36)
 \end{quote}

Our results matched findings of Khan et al.~\cite{khan_usability_2015} as well as Crawford and Renaud~\cite{crawford_understanding_2014} related to the fact that more interruptions for authentication were perceived as more annoying. They confirm findings of Reese et al.~\cite{reese_usability_2019} and Acemyan et al.~\cite{acemyan_2fa_2018} regarding that code-based 2FA received SUS scores lower than or equal 80. In relation to Reese et al. we can also confirm that the code-based 2FA SUS scores are lower than those of password-only authentication. The results also reflect findings of Zimmermann and Gerber~\cite{zimmermann_password_2020} regarding that password-only authentication received high ratings in terms of usability.%

All participants had to enter their login credentials in every task, including those of PW-ONLY. Since there was no additional security measure in this condition, PW-ONLY participants did not understand why they had to enter the credentials every time. This explains the slightly increased, but not significant, ratings for annoying (U1a) and the lower outliers in the SUS scores (U2).%

\subsubsection{Context-based User Acceptance (U3)} \label{subsection:context_based_user_acceptance}

Participants of the RBA and 2FA conditions rated their willingness to use their login procedure if they had to 
\begin{enumerate*}[label=(\roman*)]
	\item provide their email address or
	\item mobile phone number, or
	\item had to install an authenticator app on their smartphone.
\end{enumerate*}
The rating was given for seven different types of websites. Based on our classification, the website types ranged from payment data (online banking, online shopping) and personal data (email provider, social network, online storage) to less personal data (video website, comment function on a news website).%

\begin{figure}[t]
	\centering
    \includegraphics[width=\linewidth]{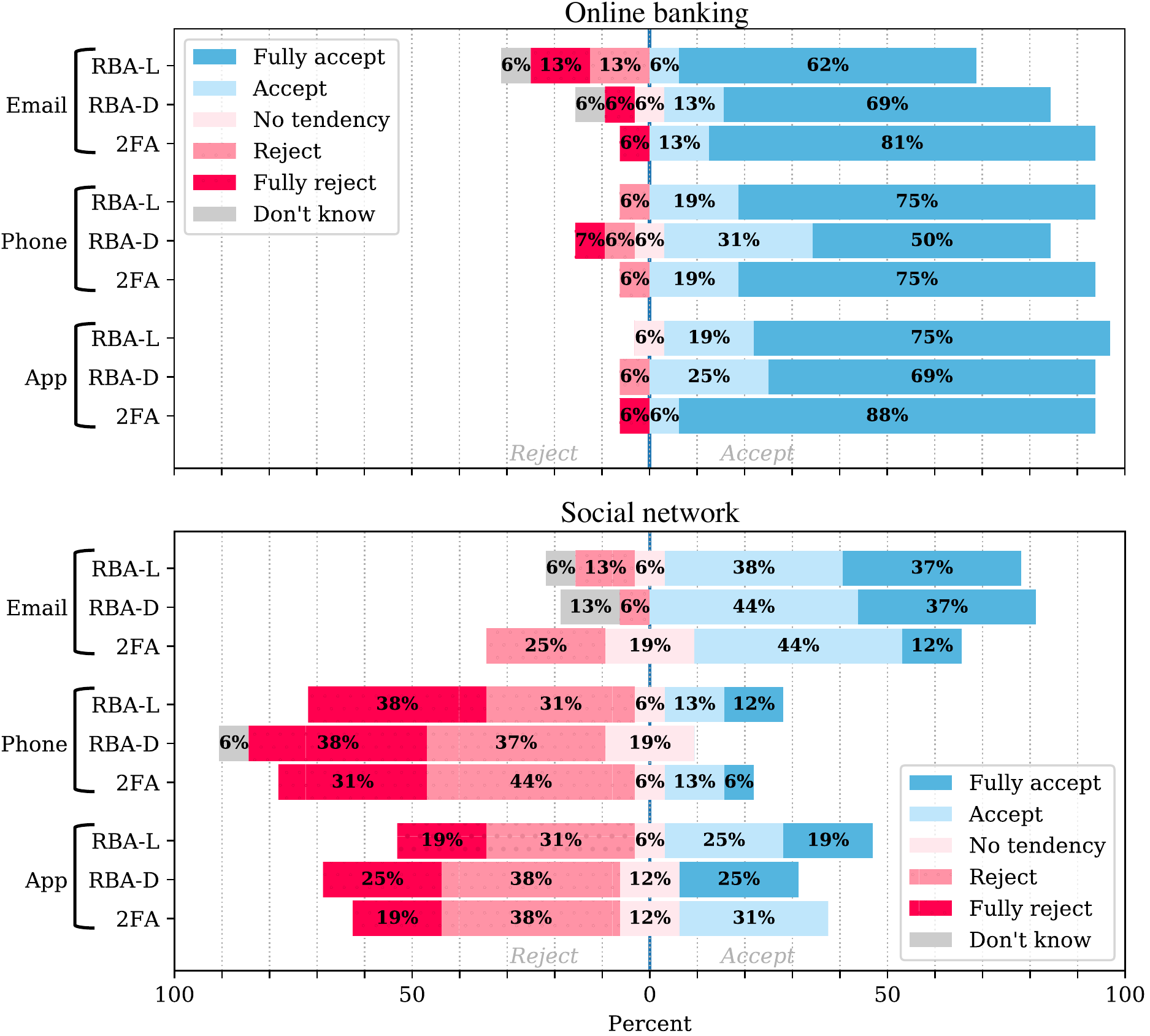}
	\caption{Context-based user acceptance (U3) responses for websites with different types of sensitive personal data involved (online banking and social network)}
	\label{fig:c_acceptance}
\end{figure}

On both RBA and 2FA conditions, the results showed a general higher acceptance for email than for mobile phone number or authenticator app%
. In the context of online banking, this general high acceptance retained for providing a mobile phone number or installing an authenticator app as well (see Figure~\ref{fig:c_acceptance}). In the following, we present an excerpt of our results. All significant results are displayed in Table~\ref{tab:dunn_c_acceptance}. Appendix~\ref{appendix:extended-results} contains all results.

\renewcommand{\arraystretch}{1.2}
\renewcommand{\arraystretch}{1.0}
\begin{table}
  \centering
  \caption{Significant (bold) K-W and Dunn-Bonferroni p-values for context-based user acceptance. We excluded p-values greater than 0.2.}
  \resizebox{\linewidth}{!}{
    \begin{tabular}{@{}rl|r|rrr@{}}
      &       & & 2FA/ & 2FA/ & RBA-L/ \\
      &       & K-W & RBA-L & RBA-D & RBA-D \\
      \midrule
    Email & Social network & \textbf{0.0457} & 0.1945 & 0.0580 & - \\
          & News website & \textbf{0.0034} & -     & \textbf{0.0029} & \textbf{0.0491} \\
          \midrule

    \\
          &       & & Email/ & Email/ & Phone/ \\
          &       & & Phone & App & App \\
          \midrule
    Online shop & RBA-LOC & \textbf{0.0137} & \textbf{0.0156} & 0.0848 & - \\
          & RBA-DEV & \textbf{0.0096} & \textbf{0.0186} & \textbf{0.0314} & - \\
    \midrule
    {Email service} & RBA-LOC & \textbf{0.0120} & \textbf{0.0091} & -     & - \\
    \midrule
    {Social network} & RBA-LOC & \textbf{0.0052} & \textbf{0.0040} & 0.1181 & - \\
          & RBA-DEV & \textbf{\textless 0.0001} & \textbf{\textless 0.0001} & \textbf{0.0123} & - \\
          & 2FA   & \textbf{0.0114} & \textbf{0.0102} & 0.1377 & - \\
    \midrule
    {Online storage} & RBA-LOC & \textbf{0.0031} & \textbf{0.0022} & -     & 0.1547 \\
          & RBA-DEV & \textbf{0.0298} & 0.0527 & 0.0763 & - \\
    \midrule
    {Video website} & RBA-LOC & \textbf{0.0038} & \textbf{0.0034} & 0.0606 & - \\
          & RBA-DEV & \textbf{0.0003} & \textbf{0.0005} & \textbf{0.0030} & - \\
          & 2FA   & \textbf{0.0072} & \textbf{0.0084} & 0.0585 & - \\
    \midrule
    News website & RBA-LOC & \textbf{0.0398} & \textbf{0.0336} & -     & - \\
          & RBA-DEV & \textbf{\textless 0.0001} & \textbf{\textless 0.0001} & \textbf{0.0015} & - \\
    \end{tabular}%
    }
  \label{tab:dunn_c_acceptance}%
\end{table}%
\renewcommand{\arraystretch}{1.2}

\textbf{Email:}
Except for the news websites, the responses showed a general acceptance for all three authentication schemes when having to provide the email address. RBA-DEV was significantly higher accepted than 2FA and RBA-LOC in the context of news website.

\textbf{Phone number:}
In all website categories, except for online banking, the acceptance to provide the phone number was significantly lower than for email to some extent. Providing the phone number was significantly less accepted than email for video website and for social network in all three conditions. 

\textbf{App:}
For RBA-DEV, the general acceptance to install an authenticator app was lower than providing the email address. Differences between app and email were significant in the contexts of video websites, news websites, online shops, and social networks. %

\paragraph{Discussion:}

The results indicate that there is a willingness to provide the mobile phone number for RBA or 2FA if very sensitive personal data or payment data is involved on a website. These results partly reflect Redmiles et al.'s~\cite{redmiles_you_2017}, Reynolds et al.'s~\cite{reynolds_tale_2018}, and Dutson et al.'s~\cite{dutson_dont_2019} observations regarding the accepted use of 2FA for only financial or sensitive data. However, personal trust in the online service seemed to be equally important, too: %

\begin{quote}
	\textit{``[I'm not providing my phone number] because [then] different websites, for example via social media, can still reach me [...]. I made experiences in the past where I was partly spammed. I received some curious messages, although I only wanted to log in in a secure way.''} (P17)
\end{quote}

Another explanation why users rejected to provide their mobile phone number on some websites was that phone numbers were regarded as more sensitive data than email addresses%
~\cite{schomakers_internet_2019}:

\begin{quote}
	\textit{``%
	If someone calls me, this is a closer contact for me than if someone writes me an email.''} (P38)
\end{quote}

Another possible factor influencing the acceptance of RBA and 2FA was the device on which the online service was mainly used. Video websites like Netflix or YouTube could be used on smart TVs as well. One participant had experiences in which the re-authentication was found annoying:

\begin{quote}
	\textit{``%
	because [...] I want to log in quickly and watch something now. On other devices I'm at the computer anyway and don't expect a problem, that I just go into the email account and get the token. As I said, on Netflix %
	[...] you do more on the TV [...] and then it's just critical.''} (P31)
\end{quote}

Additionally, for accepting RBA or 2FA on a website, users seemingly expected a certain value to be protected (e.g., access to personal data, identity theft protection). This explains why the majority of participants rejected RBA and 2FA for the comment function on a news website.%

\subsubsection{Understanding Re-Authentication (U4)}

Participants of RBA and 2FA conditions rated whether or not they understood the re-authentication. The large majority of all participants understood the re-authentication%
.%

\paragraph{Discussion:} Most of the RBA participants (RBA-LOC: 13/16, RBA-DEV: 15/16) mentioned in the semi-structured interview that this re-authentication step came after something in the behavior had changed, i.e., device or location. %
These results support the thesis, that the majority of all participants understood the sporadic re-authentication and associated it with changing situational settings.

\subsection{Security Perceptions}

In this section we evaluate and compare the security perception and perceived level of protection of the studied RBA, 2FA, and password-only authentication variants. We also identify contexts in which users feel adequately protected by RBA.

\subsubsection{Security Perception (S1)} \label{subsection:security_perception}

All participants rated the overall security of their authentication method in the exit survey. The results (see Figure~\ref{fig:overall_security}) show that participants of RBA and 2FA rated their authentication method significantly more secure than those of PW-ONLY (RBA-LOC/PW-ONLY: p=0.0013, RBA-DEV/PW-ONLY: p=0.0017, 2FA/PW-ONLY: p=0.0002). The differences between 2FA and both RBA conditions were not significant. Concluding the results, the security perception of RBA, if triggered, is significantly higher than password-only authentication and comparable to 2FA.

\paragraph{Discussion:} Participants of the two RBA conditions considered their respective authentication method as secure, since they assumed that attackers would need access to personal devices or their email accounts for a successful login:%

\begin{quote}
	\textit{``I assume that [strangers] have no access to devices on which I have already confirmed my identity. That's why I think the security is quite good''} (P6)
\end{quote}

\begin{quote}
	\textit{``[Unknown persons] don't really have a chance to access my computer or my mobile phone. Therefore, actually no chance to get the security code. [They] Must have hacked my email account somehow.''} (P13)
\end{quote}

\subsubsection{Level of Protection (S2)} \label{subsection:level_of_protection}

\begin{figure}
	\centering
	\includegraphics[width=0.94\linewidth]{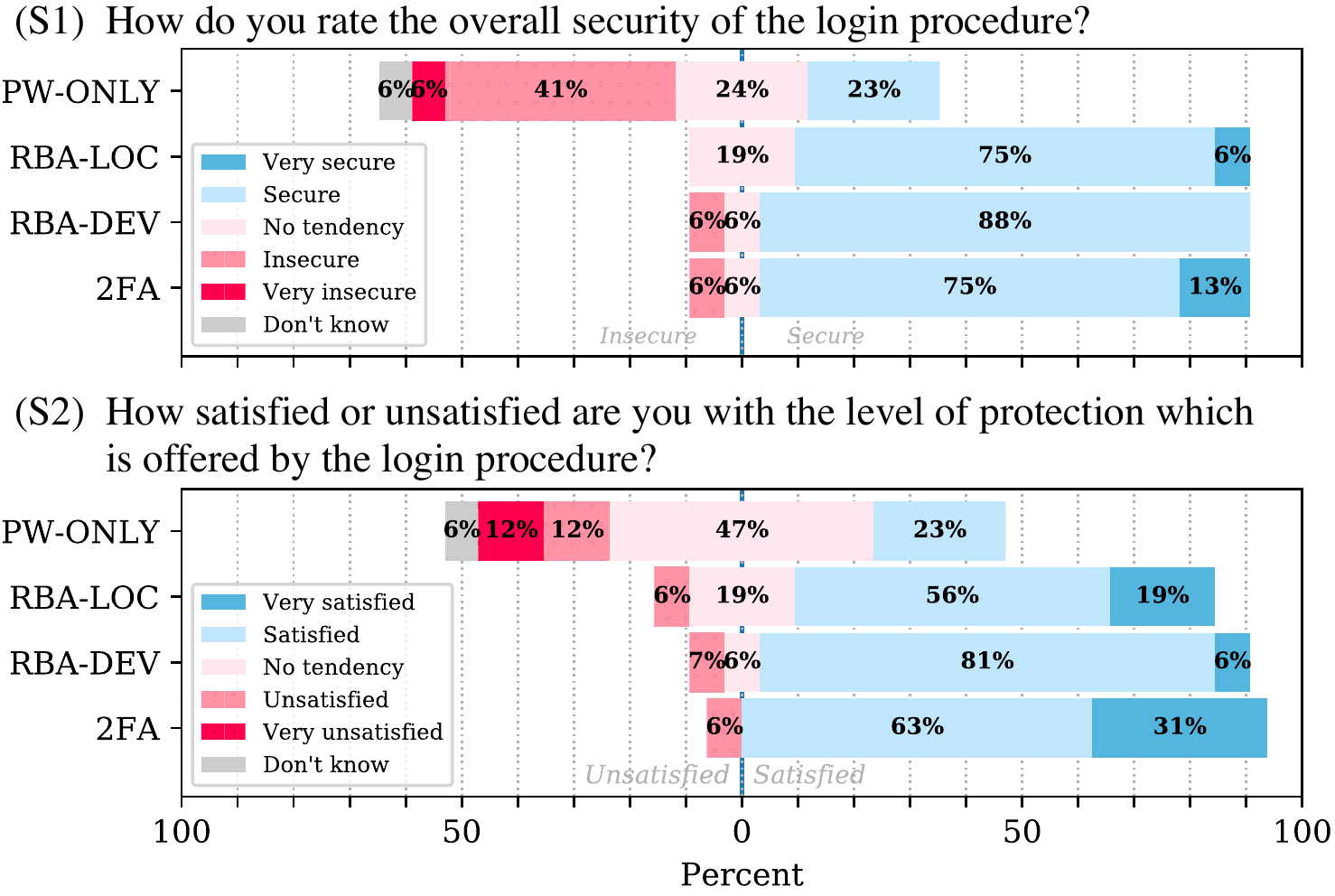}
	\caption{Participant responses for security perception (S1) and level of protection (S2). PW-ONLY participants gave significantly lower ratings than those of the other conditions.}
	\label{fig:protection_general}
	\label{fig:overall_security}
	\vspace{-1em}
\end{figure}

Participants rated how they perceived the level of protection offered by the corresponding authentication method (see Figure~\ref{fig:protection_general}). Participants of RBA and 2FA conditions were significantly more satisfied with the level of protection compared to those of the PW-ONLY condition (RBA-LOC/PW-ONLY: p=0.0126, RBA-DEV/PW-ONLY: p=0.0113, 2FA/PW-ONLY: p\textless 0.0001). There were no significant differences between 2FA and both RBA conditions. %
In conclusion, participants felt significantly more protected with RBA and 2FA than with password-only authentication. Also, RBA is comparable to 2FA regarding the perceived level of protection.

\paragraph{Discussion:}

We assume that the re-authentication played a major role for the high sense of protection. When getting into detail, all of the 2FA and RBA participants named the re-authentication as the reason for feeling protected. %
Examples:

\begin{quote}
	\textit{``The confirmation with the email address makes it feel safer.[...] %
	especially [when] it is checked again on other PCs and [one] cannot log in directly with the password and login name. And still kept simple, I found.''} (P27)
\end{quote}

\begin{quote}
	\textit{``[I felt] very secure. [...] Especially because of this confirmation email (laughs), which actually annoys you, but by doing so [...] you notice that somehow maybe a bit more security is set.''} (P39)
\end{quote}

We conclude that RBA has to be visible to users to increase security perceptions compared to password-only authentication.

\subsubsection{Context-based Level of Protection (S3)}\label{subsection:contextbased_protection}

All participants rated their satisfaction with the level of protection if the corresponding authentication method would be provided in the same manner on seven different types of websites. The websites types were identical to those mentioned in the questions for context-based user acceptance (U3).
Participants of the RBA-LOC and 2FA conditions in the online shop context, RBA-LOC participants in the social network, and 2FA participants in the online banking context showed significantly higher satisfaction with the level of protection than those of PW-ONLY.

\paragraph{Discussion:}

Online banking and online shopping involves sensitive financial data. For this reason, participants had higher demands on security than on usability in this context, as some 2FA participants noted:

\begin{quote}
	\textit{``With regard to things where financial resources come in, for example in online banking [...] or online shopping, I think it's quite good%
	. [...] %
	With things like social networking, I don't think that's absolutely necessary if you have to enter two passwords every time you log in. That would be very cumbersome.''} (P10)
\end{quote}

\begin{quote}
    \textit{``[I found it] cumbersome, but with such sensitive stuff as online banking, it's definitely justified. These aren't applications where it's about `I need one minute or three minutes', so you better take the three minutes and then you're secured.''} (P57)
\end{quote}

Besides that, we consider RBA to be suitable for contexts which involve personal data, but with lower sensitivity than online banking. Especially in these contexts, RBA outweighs password-only authentication in terms of satisfaction with the level of protection.

\section{Further Observations} \label{section:discussion}

Below, we discuss general issues which we discovered during the study but were not part of a specific research question.

\subsection{Smartphone Usage}

28 of 48 participants of the 2FA and RBA conditions used their smartphone to open the email containing the authentication code (RBA-LOC: 7, RBA-DEV: 10, 2FA: 11). %
We assume that the usage of smartphones increased the usability for re-authentication via email, as participants noted this as well:

\begin{quote}
	\textit{``I actually find it quite pleasant, because I can just read the email on my smartphone. I can solve it right away by simply taking my smartphone out of my pocket, opening the email and then entering the code.''} (P5)
\end{quote}

\begin{quote}
	\textit{``%
	[when having your smartphone] it works quite well. In the ideal case you have your smartphone with you, where you can get the email right away.''} (P6)
\end{quote}

\subsection{The Deadlock Problem}

In the RBA and 2FA conditions, re-authentication was requested (and for RBA: for the first time after the room change). Following that, participants had to log into their personal email account to get the authentication code. Participants using Gmail as their email provider (which also uses RBA) perceived deadlocks when logging in: Gmail asked for re-authentication via smartphone. Seven of 32 RBA participants and three of 16 2FA participants left their smartphones in \emph{room A} since they did not expect this re-authentication being requested%
. Thus, they were unable to %
access the %
authentication code, unless they got their smartphones back from \emph{room~A}.

We had the impression that this deadlock resulted in a frustrating user experience. %
When users perceive such a security measure as a barrier, we assume that these are likely to disable the re-authentication, if possible~\cite{crawford_understanding_2014}:
\begin{quote}
	\textit{``Sometimes [it's] very annoying, especially when the battery is flat and you don't have another device that you can log in to confirm this.%
	''} (P22)
\end{quote}

\begin{quote}
	\textit{``On Google I was very annoyed [...]%
	, because it was a shared account and I had to find someone who [...] %
	can tell me this security code.''} (P6)
\end{quote}

Resolving this deadlock problem while maintaining security for user accounts is a complex task. Especially when the email provider uses RBA as well, users will not be able to access their accounts.
Possible solutions to manage this problem can be:
\begin{enumerate*}[label=(\roman*)]
	\item More transparency by informing users of RBA and that users might be asked for re-authentication in some occasions. %
	\item Providing an alternative authentication method which could be solved without a second device or email account%
	.
	\item Allow users to define ``green zones''. As an example, we assume that a user knows about an upcoming journey to another country. Then, the user could inform the RBA-instrumented service about these specific travel circumstances.%
\end{enumerate*}

\section{Limitations}\label{limitations}

The results are limited to the persons who were willing to participate in the study. Also, the sample represents only a part of the population of a certain country. Based on the recruiting, the results are limited for young adult persons with academic education. We sampled in a country where the population is legally obliged to use 2FA for online banking and e-government. Thus, our results are applicable for societies that are used to daily 2FA use. To ensure that this is true for our sample, we asked for prior 2FA experiences in the semi-structured interview (14/16 2FA participants stated they had). %
We can not exclude that some results might differ in other countries, especially the results influenced by privacy views~\cite{schomakers_internet_2019}.

The results are also limited to websites with sensitive data involved. With a lack of sensitive data, we expect that participants would likely reject re-authentication.%

Although we put a lot of efforts into simulating a real world scenario, differences to real world usage are still there. Depending on the user behavior, RBA-based requests for re-authentication can occur less frequently in daily life than in the lab study. Concluding that, it is possible that our results regarding the RBA conditions were more negative than under real world usage. Also, the event triggering RBA was static in our study. In real world applications, however, it is possible that the false-positive rate might affect the RBA user experience negatively. We expect that RBA is not triggered when browser cookies are retained%
~\cite{hurkala_architecture_2014}. However, deleting cookies is a common user activity~\cite{rainie_anonymity_2013,varley_consumer_2018}, causing RBA to be active. Thus, we assume that the user perception is critical for RBA, especially when traveling and accessing online services abroad.

We designed the tasks with the primary goal to allow fair comparisons of RBA's and 2FA's user perceptions. 2FA using another second factor, e.g., biometrics, may offer better usability, but the same would also apply to RBA using the same biometric re-authentication scheme. The number of re-authentication steps remains the same, regardless of the re-authentication factor. %
Also, some 2FA solutions provide a %
``remember me'' option that deactivates requesting the second factor, or even both factors, for a specific time, e.g., 30 days, bringing 2FA’s look-and-feel closer to password-only authentication~\cite{duo_security_using_2019, google_google_2020, facebook_what_2020}. We see the fact that some services offer this option as an indicator that users are annoyed by frequent re-authentication~\cite{colnago_its_2018}. Again, for comparison and fairness reasons, we chose not to include a ``remember me'' function for all authentication schemes studied.

With our study, we aimed at capturing the users' understanding and perception of the targeted authentication methods. Since RBA re-authentication is commonly triggered by location or device changes, we introduced appropriate actions to support our participants in their immersion. These actions only serve as a surrounding setting and the re-authentication was designed as a secondary task. Also, our collected data relates only to the understanding and perception of the authentication methods, and not to security-critical behavior, which is potentially biased by role play (e.g., the password strength). Thus, we are convinced of a minimal role-playing bias.

Participants brought their own laptop to the study to create a realistic use case scenario. However, RBA's re-authentication requests came only under controlled conditions inside \emph{room B}. For PW-ONLY and 2FA, the login conditions did not change between the tasks. Following that, we assume that using the private laptop did not affect the experiencing of the different conditions.

\section{Related Work}\label{related-work}
We introduced a new technique with the room change, which has not been known in usable security studies to the best of our knowledge. Also, no public study evaluating the usability and security perception of RBA is known in the peer-reviewed literature to date. Koved~\cite{koved_usable_2015} investigated risk perceptions in sensitive mobile transactions by surveying participants with mock-up dialogs, which could possibly also be used in RBA systems. However, the study report has not undergone peer review, and lacks methodology, demographics as well as limitations, making the validity of the presented results difficult to judge. Nevertheless, we mention the study for the sake of completeness.

On the contrary, there are more studies evaluating usability aspects of 2FA and IA. We review them in the subsections below.

\subsection{Usability of 2FA}
Gunson et al. \cite{gunson_user_2011} compared the usability of single-factor authentication and 2FA in the context of automated telephone banking. %
The single-factor authentication %
was significantly higher rated in terms of ease of use and convenience while the 2FA approach %
was rated significantly more secure.
De Cristofaro et al. \cite{de_cristofaro_comparative_2014} compared the usability of three popular 2FA solutions with an online study%
. Their results showed an overall high usability for 2FA. As a possible explanation, they argued that the participants were not required to provide the second factor very often. %
However, we assume this authentication method to be RBA rather than 2FA. Therefore, it remains unclear if participants knew the differences between 2FA and RBA during evaluation. For this very reason, we did a direct comparison between the usability of 2FA and RBA%
.

Das et al.~\cite{das_why_2018} evaluated 2FA usability using the Yubico security key with participants. The security key aims at low-tech users with interest in securing their online services' user accounts. Though they discovered an increase in usability with the key, this did not result in increased acceptability. Our study results showed that RBA increased both usability and acceptability compared to 2FA.

Colnago et al.~\cite{colnago_its_2018} studied 2FA adoption at a university. They found that the majority of users found 2FA more pleasant to use when they reduced the number of requested re-authentication requests by activating the ``remember me'' function. Our study results showed that the user acceptance increased with fewer re-authentication requests, which was the case with RBA.

\subsection{Usability of Implicit Authentication (IA)}

Crawford and Renaud \cite{crawford_understanding_2014} evaluated user
perceptions of IA %
on mobile devices%
. The results indicated that users deactivated IA if they were asked to re-authenticate too often.
Khan et al. \cite{khan_usability_2015} conducted a two-part study to
gain insights into the usability and security perception of IA schemes.
The results showed that participants felt more secure with activated IA. In contrast to our study, both studies only simulated the authentication scheme.
Agarwal et al. \cite{agarwal_ask_2016} evaluated four different
configurations of explicit authentication schemes inside IA with a
within-group field study involving students of their university. %
Similar to our study, users preferred different authentication methods in different use case scenarios.

\section{Conclusion%
} \label{section:conclusion}

RBA is getting more and more important for website owners and website users due to increased security risks such as password database leaks, intelligent password guessing and credential stuffing attacks. The importance also increases since RBA is recommended by NIST~\cite{grassi_digital_2017} and has the potential to increase security for password-based authentication without degrading usability. To investigate this potential, we conducted a %
study with 65 participants to compare usability and security perceptions of RBA, 2FA, and password-only authentication.

Our study results provide first empirical evidence that RBA is perceived as more secure than password-only authentication and more usable than equivalent 2FA variants. 
We found that the user acceptance of RBA is dependent on the type of website and the device on which it is mainly used. In general, RBA using email address confirmation is accepted for websites which store a certain amount of sensitive data. In contrast to that, RBA using mobile phone numbers or authenticator apps for re-authentication is less accepted.
Thus, deploying RBA has to be considered carefully for each use case scenario. Special attention has also to be taken if access to the re-authentication factor (e.g., email address) is protected with RBA as well, since this could result in locking out users.

Regarding the security perceptions, our results suggest that RBA is considered to be comparably secure as 2FA for a wide range of websites. Only for high security demands, such as posed by online banking, 2FA is preferable over RBA, due to the higher feeling of protection in this context.

Our results indicate that users have a demand for strong security on websites, especially when sensitive data is involved. In contrast to 2FA, RBA can provide this security with minimal burden on the user~\cite{wiefling_evaluation_2020}. This is probably one of the reasons why users preferred RBA over 2FA in our study and why 2FA has low adoption rates in the wild~\cite{milka_anatomy_2018}. %
Hence, almost all websites involving sensitive data should consider deploying RBA to protect their users.

\begin{acks}
Thanks to all participants for their voluntary participation in the study. We thank all anonymous reviewers for their constructive feedback, which greatly contributed to improve the paper. Many thanks to Michael Kloos for his support in conducting the study. In the same way, we thank Jan Tolsdorf, Paul Höller, Peter Leo Gorski, and Tanvi Patil for their support, including study coding, proof reading, and providing feedback on drafts of the paper. We would also like to thank all the staff, organizations, and people at the universities who supported us to recruit the participants, including among others in alphabetical order Annette Ricke, Bettina Menden, Hoai Viet Nguyen, Jan Herrmann, Jeanette Dietz, Kölncampus, Marc Kastner, Mendy Stoll, Michael Lorth, Thomas Krupp, Tobias Mengel, and Zelal Ates. This research was supported by the research training group ``Human Centered Systems Security'' (NERD.NRW) sponsored by the state of North Rhine-Westphalia.
\end{acks}

\vspace{\fill}

	\bibliographystyle{ACM-Reference-Format}
	\bibliography{literature}

	\appendix

\section{Website}

\subsection{Additional Authentication Dialog} \label{appendix:dialog}

Based on six re-authentication dialogs, we created a generic dialog representing RBA state-of-the-art deployments (see Section~\ref{design-decisions}). We categorized the wording and design decisions inside the corresponding dialogs. The wordings and design decisions with the highest occurrences were selected for the final dialog (see Figure~\ref{fig:auth-rba}).

\renewcommand{\arraystretch}{1.3}
\renewcommand{\arraystretch}{1.0}
\begin{table}[H]
  \centering
  \caption{Ranking the wording and design decisions for the RBA identity confirmation dialog of our study website. Bold highlighted: Taken for the final dialog.}
  \resizebox{0.7\linewidth}{!}{
  \begin{threeparttable}
  	\begin{tabular}{@{}r|ccccccc@{}}
  		
  		& \rot{Amazon} & \rot{Facebook} & \rot{GOG.com} & \rot{Google} & \rot{LinkedIn} & \rot{Microsoft} \\
  		\midrule
  		\textit{Process} &       &       &       &       &       &  \\
  		\textbf{Identity/Login Verify} & \Circle & \CIRCLE     &  \Circle & \CIRCLE &  \Circle & \CIRCLE \\
  		Identity/Login Check & \CIRCLE & \Circle &  \Circle  &  \Circle  & \CIRCLE     & \Circle \\
  		Two-Step & \Circle &  \Circle  &   \CIRCLE    &  \Circle  & \Circle & \Circle  \\
  		\midrule
  		\textit{Additional Factor} &       &       &       &       &       &  \\
 		\textbf{Security Code} & \Circle & \CIRCLE     & \CIRCLE & \CIRCLE &  \Circle  &  \Circle \\
  		Verification Code &  \CIRCLE  &  \Circle & \Circle & \Circle  & \LEFTcircle     &  \LEFTcircle \\
  		\midrule
  		\textit{Email Address Display} &       &       &       &       &       &  \\
  		\textbf{Censored/Not shown} & \CIRCLE & \CIRCLE &  \Circle &  \Circle & \LEFTcircle\tnote{D} & \CIRCLE \\
  		Uncensored &  \Circle &  \Circle & \CIRCLE & \CIRCLE & \LEFTcircle\tnote{M} &  \Circle \\
  		\midrule
  		\textit{Authentication code} &       &       &       &       &       &  \\
  		\textbf{Six digits} & \CIRCLE & \CIRCLE & \Circle & \CIRCLE & \CIRCLE & \Circle \\
  		Seven digits & \Circle & \Circle & \Circle & \Circle & \Circle & \CIRCLE \\
  		Four digits & \Circle & \Circle & \CIRCLE & \Circle & \Circle & \Circle \\
  		\bottomrule
  	\end{tabular}%
	\begin{tablenotes}
		\item[\CIRCLE] Present
		\item[\Circle] Not present
		\item[\LEFTcircle] Not in all dialogs\\
		\item[D] Desktop view only
	  	\item[M] Mobile view only
	\end{tablenotes}
  \end{threeparttable}
}
  \label{tab:dialogranking}
\end{table}%
\renewcommand{\arraystretch}{1.3}

\subsection{Additional Authentication Email} \label{appendix:email}

For the RBA and 2FA conditions, the study website sent an email to participants to confirm the claimed identity. The email content is based on additional authentication emails of six online services (see Section~\ref{design-decisions}).

\begin{figure}[H]
	\fbox{%
		\parbox{\linewidth}{%
			{\small 
				\texttt{From:\textit{ [website]} Security}\newline
				\texttt{Subject: Your personal security code}
				\bumper
				Dear \textit{[website]} user,\newline
				someone just tried to sign in to your \textit{[website]} account.\newline
				
				If you were prompted for a security code, please enter the following\newline to complete your sign-in:\newline
				
				\textit{[Six-digit authentication code]}\newline
				
				If you were not prompted, please change your password \newline immediately in the profile settings of \textit{[website]}.\newline
				
				Thanks, the \textit{[website]} team}}
	}
	\caption{Email, which was sent to the participants%
	}
	\label{fig:email-text}
\end{figure}

\section{Study Tasks}\label{appendix:study-tasks}

\normalsize
All tasks\censorchange{}{\footnote{We changed some names and wordings inside the tasks for blind review.}} were printed one after the other on paper. Participants were asked to turn to the next sheet containing the next task when the task was completed.

\small
\subsection*{\small Task 1}\label{appendix:study-tasks-1}

\begin{enumerate}
	\setlength\itemsep{0em}
	\item Turn on your personal laptop.
	\item Open a web browser of your choice.
	\item Register on the cloud storage website [website].
	
	The website is available at:\newline
	\texttt{https://[website]/register}
	
	Your access code for the registration is: [Access code]
	
	For the registration, please use your private email address and a password.
	
	\item Log out after registration.
\end{enumerate}

\subsection*{\small Task 2}

You're about to have a business meeting with a potential client. You're going to give a talk there.

\begin{enumerate}
	\setlength\itemsep{0em}
	\item Log into [website].
	\item Upload the presentation and the minutes for this meeting there.
	
	Both files are stored on a USB flash drive, which is located in front of you.
	
	\item Log out afterwards.
\end{enumerate}

When you've finished this task, please call the study conductor by pressing the grey button.

\begin{center}
\bumper
\vspace{1em}
\textit{Participant is brought from \emph{room A} to \emph{room B} (the ``internet cafe'').}
\bumper
\end{center}

\subsection*{\small Task 3}

You are on your way to the customer. Shortly before you reach your destination, you noticed that you have forgotten your laptop with the presentation and important data.

Luckily, there is an open internet cafe next to the customers destination so that you can access your data there.

You are now inside the internet cafe on a computer assigned to you. You've bought a USB flash drive beforehand, which is now located in front of you.

\begin{enumerate}
	\setlength\itemsep{0em}
	\item Open the Chrome browser.
	\item Log into [website].
	\item Download the presentation (not the meeting minutes) for your talk there.
	\item Log out afterwards.
	\item Save the presentation on the USB flash drive so that you can open it later on the customer's presentation computer.
\end{enumerate}

\subsection*{\small Task 4}

Mr. \censorchange{Berner}{Johnson}, a business partner of the \censorchange{CLOUST}{[Company name]}\censorchange{ AG}{, Inc.} gets in touch with you. He wants to know the quarterly figures (2nd quarter of 2018) of the business report. You can access the business report via [website].

\begin{enumerate}
	\setlength\itemsep{0em}
	\item Log into [website].
	\item Look for the current quarterly figure (revenue from 2nd quarter of 2018) from the business report and send him the figure via email.
	
	His email address is \censorchange{berner}{johnson}@[Company domain name].\censorchange{de}{com}\footnote{We bought an internet domain representing this fictional company (not linked to our university) and controlled the email address of the business partner.}.
	\item Don't forget to log out afterwards since you're going to give the talk afterwards and therefore have to leave the internet cafe.
\end{enumerate}

\subsection*{\small Task 5}

Your colleague \censorchange{Alisa Berger}{Lisa Smith} gets in touch with you. She needs a photo of you to introduce the company to another customer.

The talk is only in 15 minutes and you remember that you can upload photos to [website] and share them with her. Also, the computer inside the internet cafe has a camera which you can use.

\begin{enumerate}
	\setlength\itemsep{0em}
	\item Log into [website].
	\item Click the button ``Take a picture''.
	\item Now take a picture of yourself there which is stored automatically on the website.
	\item Share this picture with \censorchange{Alisa Berger}{Lisa Smith}.
	\item Log out afterwards.
\end{enumerate}

\subsection*{\small Task 6}

You gave the talk. After a short break, a meeting with the potential customer should take place.

For a good preparation, you have already stored the meeting minutes on [website]. You have borrowed a tablet PC for the meeting, which is connected to the customer's public Wi-Fi network.

\begin{enumerate}
	\setlength\itemsep{0em}
	\item Get the tablet PC from the right drawer of the desk.
	\item Open the Chrome browser on the tablet PC.
	\item Log into [website] with the tablet PC.
	\item Open the meeting minutes.
\end{enumerate}

Afterwards, please call the study conductor by pressing the grey button.

\begin{center}
\bumper
\vspace{1em}
\textit{Participant is brought from \emph{room B} to \emph{room A} again.}
\bumper
\end{center}

\subsection*{\small Task 7}

You're at home again where you've found your forgotten laptop. Now you want to delete the data which you no longer need on [website].

\begin{enumerate}
	\setlength\itemsep{0em}
	\item Turn on your laptop.
	\item Open a web browser of your choice.
	\item Log into [website] with your laptop.
	\item Delete the presentation, the meeting minutes, and the picture you uploaded.
	\item Delete your user profile via the ``Profile'' menu tab.
\end{enumerate}

Afterwards, please call the study conductor by pressing the grey button.

\section{Questions} \label{appendix:questions}

\subsection{Modified SUS Surveys}\label{appendix:sus}

\normalsize

Participants responded on a five-point Likert scale (5 - Strongly agree, 1 - Strongly disagree). The scale direction varied for a randomly selected half of participants in each study group. The question order varied randomly for each participant.

\small
\subsubsection{SUS 1: Website}

\begin{itemize}[leftmargin=2em, itemsep=0em]
	\item I think that I would like to use this website frequently
	\item I found the website unnecessarily
complex
	\item I thought the website was easy
to use
	\item I think that I would need the
support of a technical person to
be able to use this website
	\item I found the various functions on
this website were well integrated	
	\item I thought there was too much
inconsistency on this website
	\item I would imagine that most people
would learn to use this website
very quickly
	\item I found the website very
cumbersome to use
	\item I felt very confident using the
website
	\item I needed to learn a lot of
things before I could get going with this website
\end{itemize}

\subsubsection{SUS 2: Login Procedure}

\begin{itemize}[leftmargin=2em, itemsep=0em]
	\item I think that I would like to use this login procedure frequently
	\item I found the login procedure unnecessarily
complex
	\item I thought the login procedure was easy
to use
	\item I think that I would need the
support of a technical person to
be able to use this login procedure
	\item I found the various functions of
this login procedure were well integrated	
	\item I thought there was too much
inconsistency of this login procedure
	\item I would imagine that most people
would learn to use this login procedure
very quickly
	\item I found the login procedure very
cumbersome to use
	\item I felt very confident using the
login procedure
	\item I needed to learn a lot of
things before I could get going with this login procedure
\end{itemize}

\subsection{Exit Survey}\label{appendix:exit-survey}

\normalsize

Participants responded on a five-point Likert scale including a ``don't know'' option. The scale direction varied for a randomly selected half of participants in each study group. The question order and the order of the subquestions varied randomly for each participant.

\small
\begin{itemize}[leftmargin=1em, topsep=1em, itemsep=0em]
\item
  How satisfied or unsatisfied are you with the level of protection
  which is offered by the login procedure?
  
  (5 - Very satisfied, 1 - Very unsatisfied)
  
\item
  How satisfied or unsatisfied are you with the level of protection
  of the the login procedure, if it is provided in the same manner on the following types
  of websites?
  
  (5 - Very satisfied, 1 - Very unsatisfied)
  
  \noindent
  \begin{answerlist}
  	\setlength\itemsep{0em}
  	\item
  	Online banking
  	\item
  	Online shop
  	\item
  	Email service
  	\item
  	Social network
  	\item
  	Online storage (Dropbox, Google Drive and others)
  	\item
  	Video-sharing website
  	\item
  	Comment function on a news website
  \end{answerlist}

\item How annoying or not annoying did you perceive this login procedure?

	(5 - Not annoying at all, 1 - Very annoying)

\item How much time does this login procedure take according to your perception?

	(5 - Very little time, 1 - Very much time)
	
\item How tiring or not-tiring did you find this login procedure?

	(5 - Not tiring at all, 1 - Very tiring)
	
\item How did you perceive the interruptions for confirming the identity?

	(5 - Not annoying at all, 1 - Very annoying)
	
\item How do you rate the overall security of the login procedure?

	(5 - Very secure, 1 - Very insecure)

\item \textit{[2FA, RBA]} Please rate your agreement with the following statement:\linebreak\textbf{I understood why I had to confirm my Identity a second time.}

(5 - Strongly agree, 1 - Strongly disagree)

\item \textit{[2FA, RBA]} How secure do you find this login procedure compared to a login procedure with password and without identity confirmation?

	(5 - Very more secure, 1 - Very more insecure)

\item \textit{[2FA, RBA]} Would you use this login procedure?

	(5 - Yes, very sure, 1 - No, definitely not)
	
\item \textit{[2FA, RBA]} How much would you accept or reject the identity confirmation on the following types of websites, if you would have to enter your email address for this purpose?

	(5 - Fully accept, 1 - Fully reject)

  \noindent
  \begin{answerlist}
    \setlength\itemsep{0em}
	\item
	Online banking
	\item
	Online shop
	\item
	Email service
	\item
	Social network
	\item
	Online storage (Dropbox, Google Drive and others)
	\item
	Video website
	\item
	Comment function on a news website
  \end{answerlist}

\item \textit{[2FA, RBA]} How much would you accept or reject the identity confirmation on the following types of websites, if you would have to enter your mobile phone number for this purpose?

(5 - Fully accept, 1 - Fully reject)

\textit{Same categories as in the question before.}

\item \textit{[2FA, RBA]} How much would you accept or reject the identity confirmation on the following types of websites, if you would have to install a special app on your smartphone for this purpose?

(5 - Fully accept, 1 - Fully reject)

\textit{Same categories as in the question before.}

\end{itemize}

\subsection{Semi-structured Interview}

\begin{enumerate}[itemsep=0em]
	\item What did you like on the website?
	\item What didn't you like on the website?
	\item What did you like on the login procedure?
	\item What didn't you like on the login procedure?
	\item Would you change anything on the login procedure?
	\item How was your security perception when you were using the website?
	\item Do you have suggestions for alternative authentication methods? \newline\textit{[If yes:]}: Which ones?
	\item \textit{[2FA, RBA]} Can you explain how this login procedure works?
	\item \textit{[2FA, RBA]} Can you trace back the identity confirmation to a certain behavior?
	\item \textit{[2FA, RBA]} Have you ever come into contact with such a login procedure?\newline\textit{[If yes:]} Where exactly? How was your perception there?
\end{enumerate}

\vspace{\fill}
\newpage
\clearpage

\normalsize
\section{Extended Results} \label{appendix:extended-results}

\begin{minipage}{1.3\linewidth}
	\begin{figure}[H]
		\centering
		\begin{subfigure}[b]{\linewidth}
			\centering
			\includegraphics[width=\textwidth]{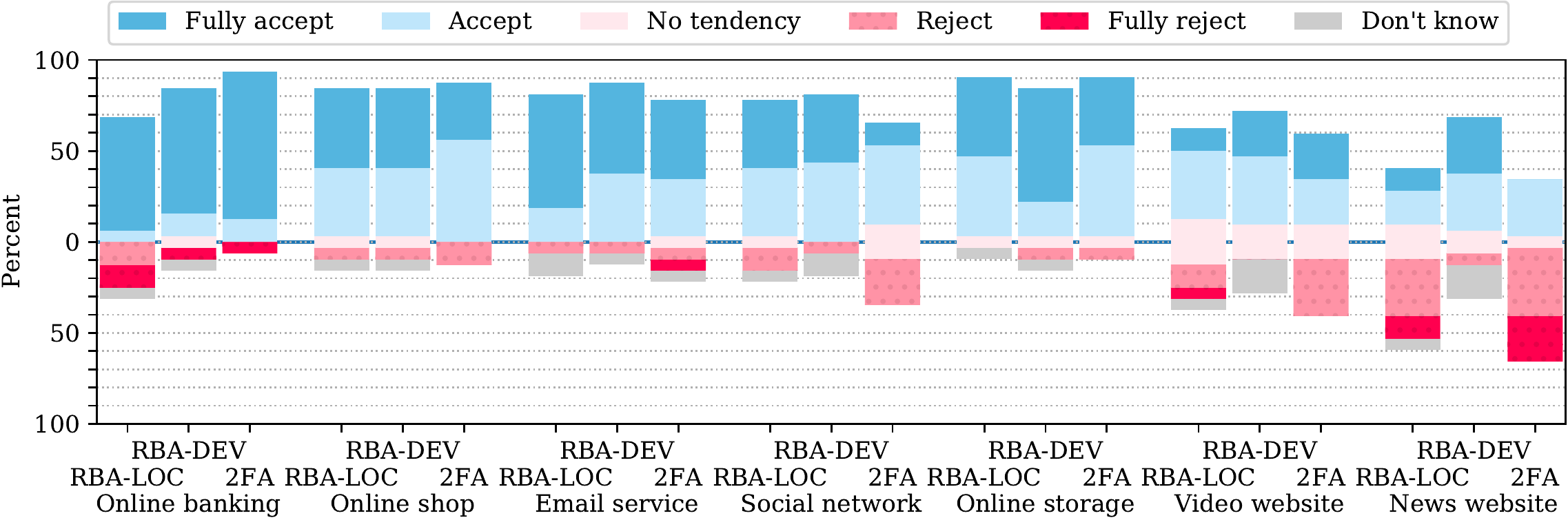}
			\caption{How much would you accept or reject the identity confirmation on the following types of websites, if you would have to enter your email address for this purpose?}
			\label{fig:c_full_acceptance_email}
		\end{subfigure}
		\begin{subfigure}[b]{\linewidth}
			\centering
			\includegraphics[width=\textwidth]{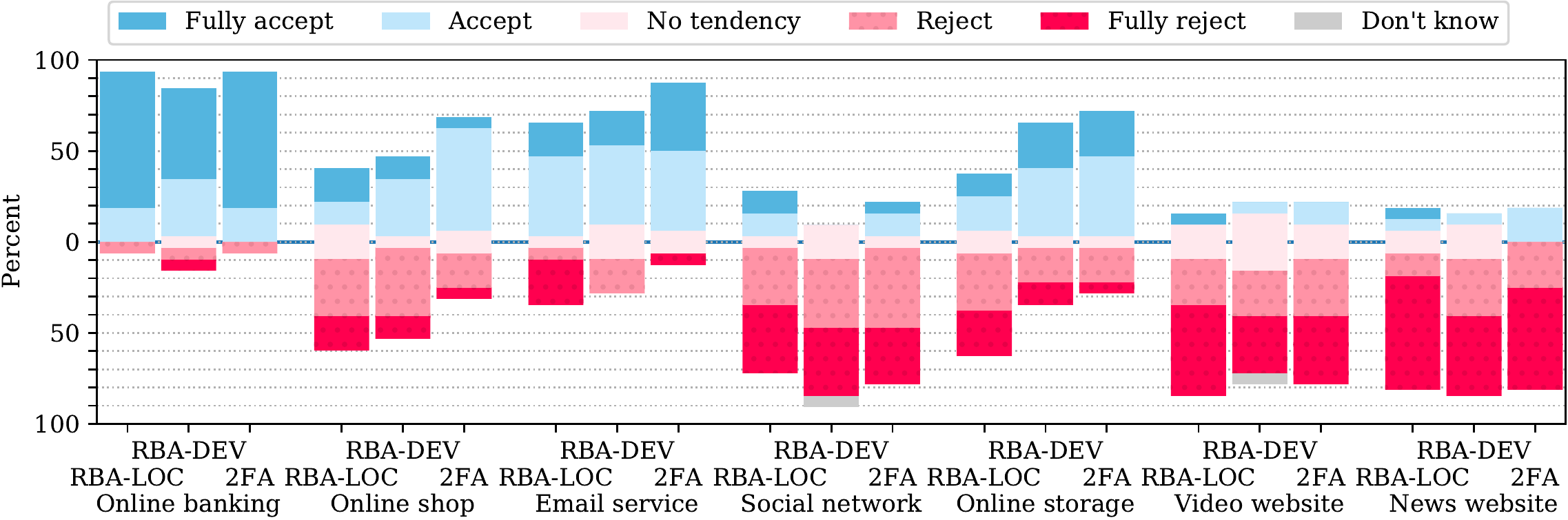}
			\caption{How much would you accept or reject the identity confirmation on the following types of websites, if you would have to enter your mobile phone number for this purpose?}
			\label{fig:c_full_acceptance_phone}
		\end{subfigure}
		\begin{subfigure}[b]{\linewidth}
			\centering
			\includegraphics[width=\textwidth]{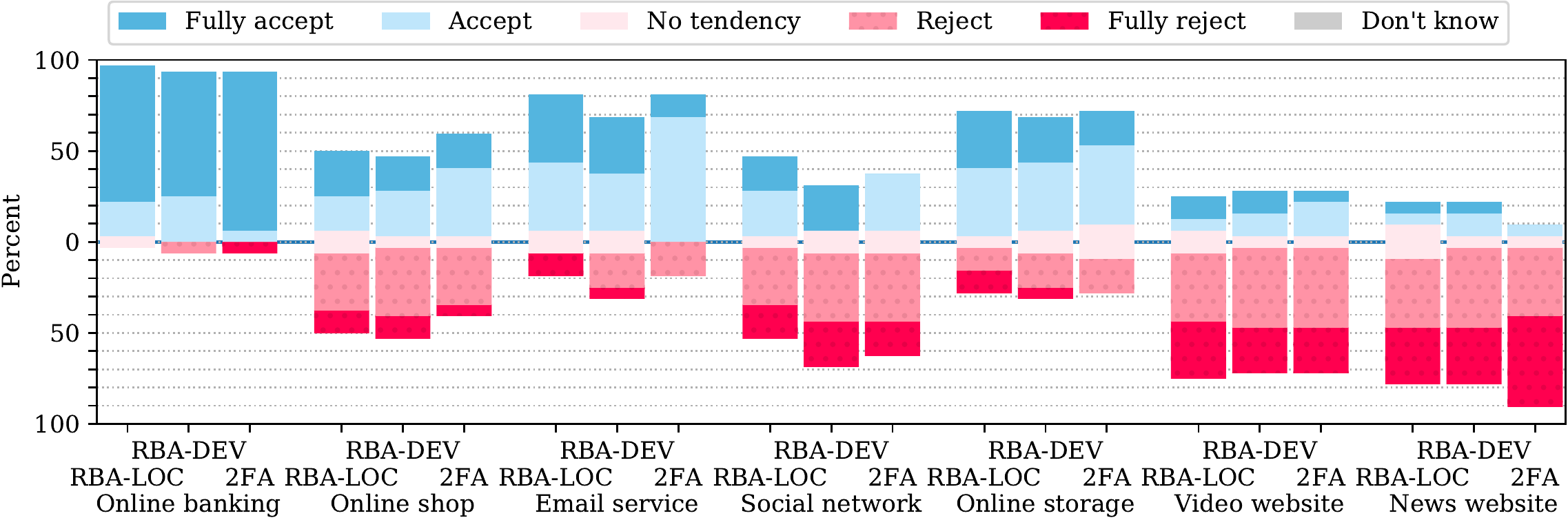}
			\caption{How much would you accept or reject the identity confirmation on the following types of websites, if you would have to install a special app on your smartphone for this purpose?}
			\label{fig:c_full_acceptance_app}
		\end{subfigure}
		\caption{Likert plots showing the responses to the context based user acceptance questions}
		\label{fig:c_full_acceptance}
	\end{figure}
	\begin{figure}[H]
		\vspace{-1em}
		\centering
		\includegraphics[width=\linewidth]{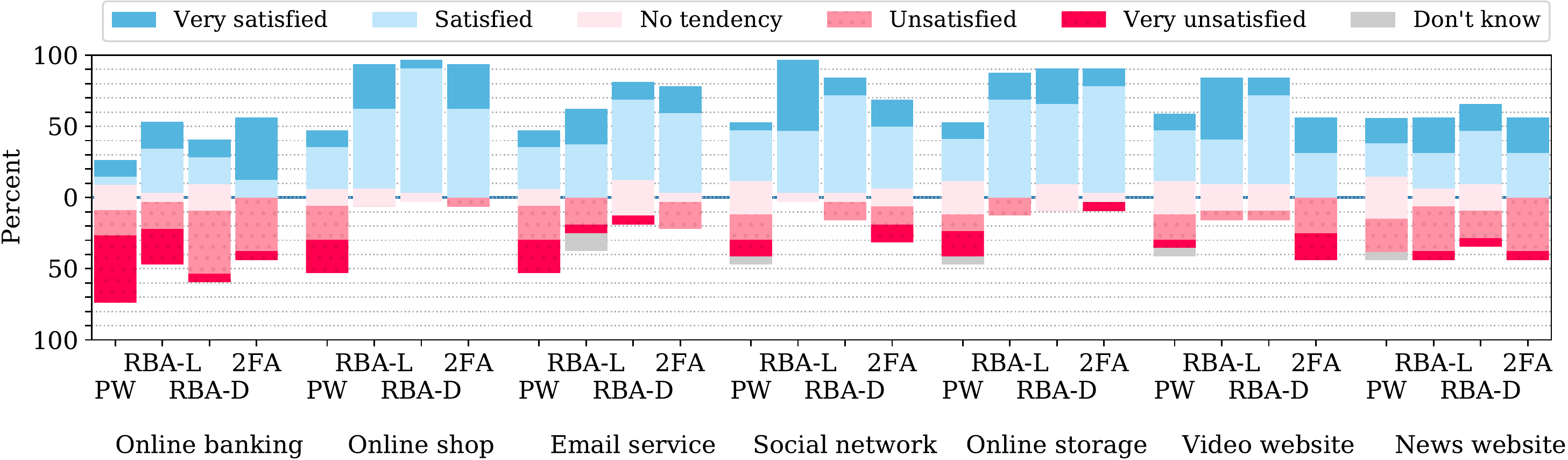}
		\caption{Likert plots showing the responses to the question ``How satisfied or unsatisfied are you with the level of protection of the the login procedure, if it is provided in the same manner on the following types of websites?''}
		\label{fig:protection_website}
	\end{figure}
\end{minipage}

\hspace{0.2\linewidth}
\begin{minipage}{0.8\linewidth}
	\begin{figure}[H]
		\centering
		\includegraphics[width=\linewidth]{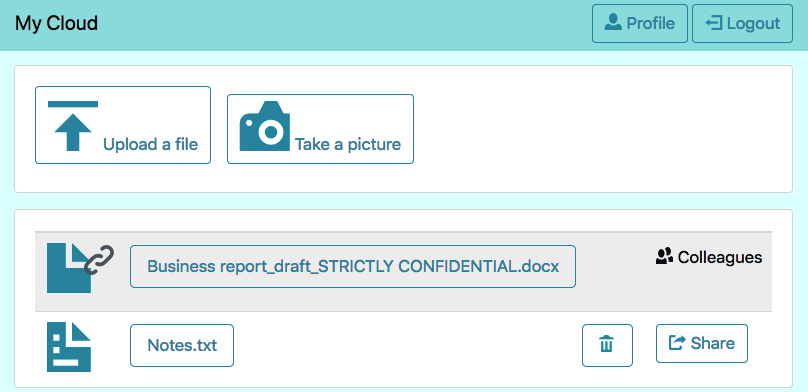}
		\caption{Desktop view of the %
			study website}
		\label{fig:study-website}
	\end{figure}

	\begin{figure}[H]
		\vspace{-1em}
		\centering
		\includegraphics[width=\linewidth]{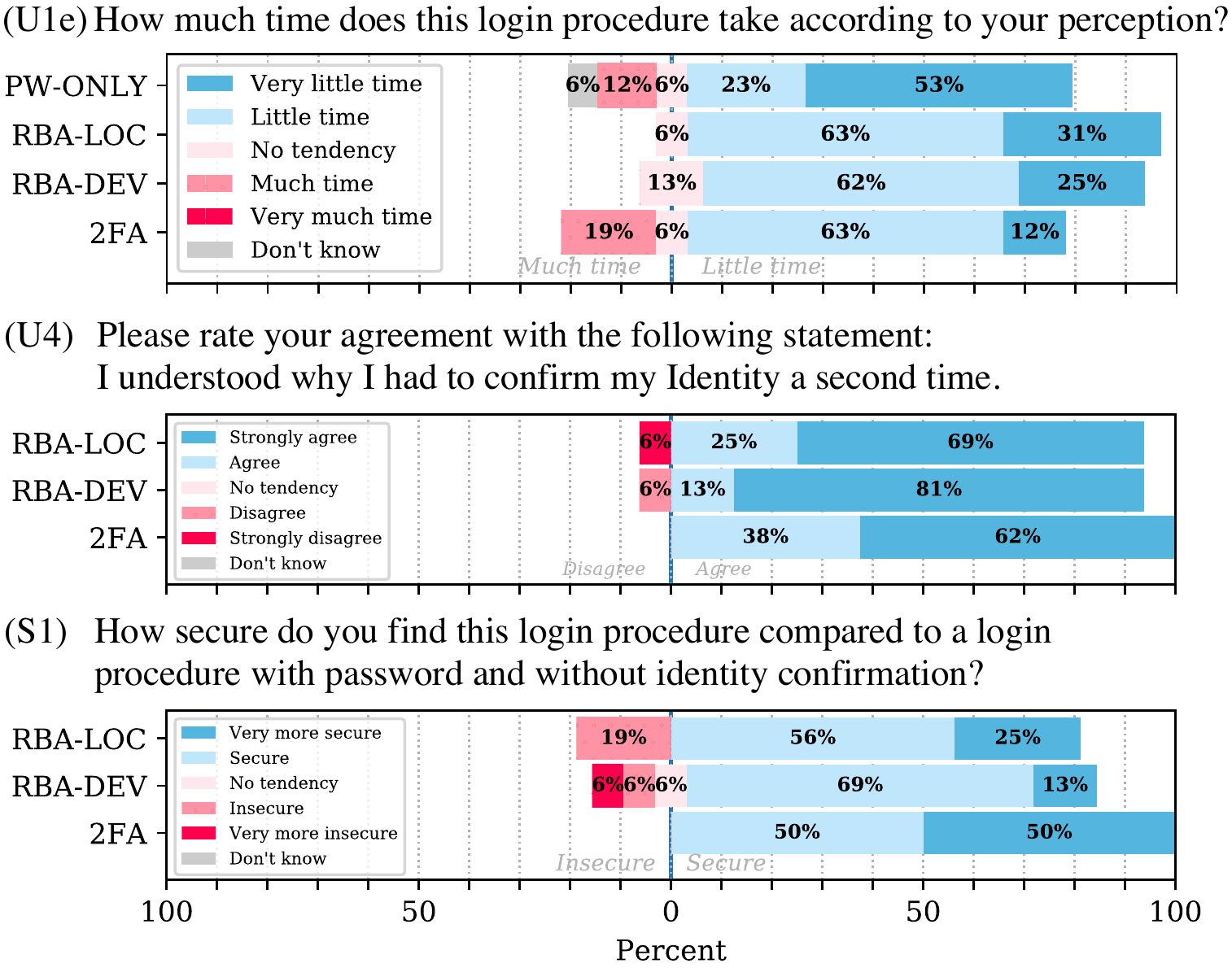}
		\caption{Additional responses to the user acceptance (U1), understanding (U4) and security perception (S1) questions}
		\label{fig:understanding_auth}
		\label{fig:security_comparedtopw}
	\end{figure}

\renewcommand{\arraystretch}{1.3}
\renewcommand{\arraystretch}{1.0}
\begin{table}[H]
	\vspace{-1em}
	\centering
	\caption{Kruskal-Wallis omnibus test and Dunn-Bonferroni post-hoc analysis (p-values) for context-based user acceptance (U3) of the identity confirmation between providing an email address or mobile phone number, or installing an authenticator app. Bold: Significant, *: 1.000}
	\resizebox{\linewidth}{!}{\begin{tabular}{rl|rr|rrr}
			&       & \multicolumn{2}{c|}{Kruskal-Wallis} & \multicolumn{3}{c}{Dunn-Bonferroni} \\
			&       &            &       & Email/ & Email/ & Phone/ \\
			&       & $\chi ^2$  & p     & Phone  & App   & App \\
			\midrule
			{Online banking} & RBA-LOC & 0.9752 & 0.6141 & *     & *     & * \\
			& RBA-DEV & 1.9903 & 0.3697 & 0.6363 & *     & 0.7059 \\
			& 2FA   & 0.6536 & 0.7212 & *     & *     & * \\
			\midrule
			{Online shop} & RBA-LOC & 8.5748 & \textbf{0.0137} & \textbf{0.0156} & {0.0848} & * \\
			& RBA-DEV & 9.2877 & \textbf{0.0096} & \textbf{0.0186} & \textbf{0.0314} & * \\
			& 2FA   & 4.5662 & 0.1020 & 0.1750 & 0.2134 & * \\
			\midrule
			{Email service} & RBA-LOC & 8.8533 & \textbf{0.0120} & \textbf{0.0091} & 0.2209 & 0.6699 \\
			& RBA-DEV & 5.4381 & {0.0659} & {0.0898} & 0.1929 & * \\
			& 2FA   & 2.1842 & 0.3355 & *     & 0.4927 & 0.7895 \\
			\midrule
			{Social network} & RBA-LOC & 10.5111 & \textbf{0.0052} & \textbf{0.0040} & 0.1181 & 0.7310 \\
			& RBA-DEV & 18.8850 & \textbf{\textless 0.0001} & \textbf{\textless 0.0001} & \textbf{0.0123} & 0.3964 \\
			& 2FA   & 8.9550 & \textbf{0.0114} & \textbf{0.0102} & 0.1377 & * \\
			\midrule
			{Online storage} & RBA-LOC & 11.5340 & \textbf{0.0031} & \textbf{0.0022} & 0.4284 & 0.1547 \\
			& RBA-DEV & 7.0275 & \textbf{0.0298} & {0.0527} & {0.0763} & * \\
			& 2FA   & 2.8976 & 0.2348 & 0.5682 & 0.3321 & * \\
			\midrule
			{Video website} & RBA-LOC & 11.1586 & \textbf{0.0038} & \textbf{0.0034} & {0.0606} & * \\
			& RBA-DEV & 16.4038 & \textbf{0.0003} & \textbf{0.0005} & \textbf{0.0030} & * \\
			& 2FA   & 9.8765 & \textbf{0.0072} & \textbf{0.0084} & {0.0585} & * \\
			\midrule
			{News website} & RBA-LOC & 6.4476 & \textbf{0.0398} & \textbf{0.0336} & 0.4764 & 0.7553 \\
			& RBA-DEV & 18.8718 & \textbf{\textless 0.0001} & \textbf{\textless 0.0001} & \textbf{0.0015} & * \\
			& 2FA   & 4.1956 & 0.1227 & 0.2334 & 0.2233 & * \\
		\end{tabular}%
	}
	\label{tab:omnibusposthoc_contextuseracceptance}%
\end{table}%
\renewcommand{\arraystretch}{1.3}
\end{minipage}

\onecolumn
\renewcommand{\arraystretch}{1.3}
\renewcommand{\arraystretch}{1.0}
\begin{table}[htbp]
  \centering
 	\caption{Results of the Kruskal-Wallis omnibus test and Dunn-Bonferroni post-hoc analysis (p-values) for the exit survey questions. Bold: Significant, *: 1.000}
  \resizebox{0.83\linewidth}{!}{
    \begin{tabular}{l|rr|rrr|rr|r}
          & \multicolumn{2}{c|}{Kruskal-Wallis} & \multicolumn{6}{c}{Dunn-Bonferroni} \\
          &       &       & \multicolumn{3}{c}{2FA} & \multicolumn{2}{|c}{PW-ONLY} & \multicolumn{1}{|c}{RBA-LOC}\\
          & \multicolumn{1}{c}{$\chi ^2$} & \multicolumn{1}{c|}{p} & \multicolumn{1}{l}{PW-ONLY} & \multicolumn{1}{l}{RBA-LOC} & \multicolumn{1}{l}{RBA-DEV} & \multicolumn{1}{|l}{RBA-LOC} & \multicolumn{1}{l}{RBA-DEV} & \multicolumn{1}{|l}{RBA-DEV} \\
    \midrule
    U1 &       &       &       &       &       &       &  \\
    \hspace{1em}General annoyance & 17.9578 & \textbf{0.0004} & {0.0560} & \textbf{0.0010} & \textbf{0.0022} & *     & *     & * \\
    \hspace{1em}Perceived time & 5.3885 & 0.1455 & 0.1505 & 0.6300 & *     & *     & *     & * \\
    \hspace{1em}Tiring & 10.7254 & \textbf{0.0133} & {0.0725} & \textbf{0.0122} & 0.3118 & *     & *     & * \\
    \hspace{1em}Re-authentication annoyance & 7.2587 & \textbf{0.0265} &       & \textbf{0.0331} & 0.1225 &       &       & * \\
    \hspace{1em}Re-authentication understand & 1.0736 & 0.5846 &       & *     & 0.9594 &       &       & * \\
    \hspace{1em}Like to use login procedure & 10.1893 & \textbf{0.0061} &       & \textbf{0.0117} & *     &       &       & \textbf{0.0260} \\
    \midrule
    U2: SUS-Score & 13.9371 & \textbf{0.0030} & \textbf{0.0093} & * & * & {0.0523} & \textbf{0.0073} & * \\
    \midrule
    U2: SUS: Login &       &       &       &       &       &       &  \\
    \hspace{1em}Use more frequently & 13.2633 & \textbf{0.0041} & 0.7633 & \textbf{0.0185} & \textbf{0.0078} & 0.8362 & 0.4914 & * \\
    \hspace{1em}Unnecessarily complex & 18.5616 & \textbf{0.0003} & \textbf{0.0005} & \textbf{0.0420} & \textbf{0.0026} & *     & *     & * \\
    \hspace{1em}Easy to use & 12.6901 & \textbf{0.0054} & \textbf{0.0034} & 0.1084 & {0.0964} & *     & *     & * \\
    \hspace{1em}Need support & 2.3167 & 0.5093 & *     & *     & *     & *     & *     & * \\
    \hspace{1em}Functions well integrated & 5.6887 & 0.1278 & *     & 0.5794 & 0.2395 & *     & 0.6447 & * \\
    \hspace{1em}Too much inconsistency & 6.0665 & 0.1084 & 0.1024 & *     & 0.7143 & 0.9288 & *     & * \\
    \hspace{1em}Quickly learn to use & 3.7299 & 0.2921 & *     & 0.5200 & *     & *     & *     & 0.6252 \\
    \hspace{1em}Cumbersome to use & 19.6675 & \textbf{0.0002} & \textbf{0.0005} & \textbf{0.0049} & \textbf{0.0027} & *     & *     & * \\
    \hspace{1em}Felt confident to use & 2.5935 & 0.4586 & *     & *     & *     & *     & *     & * \\
    \hspace{1em}Need to learn a lot & 1.8281 & 0.6088 & *     & *     & *     & *     & *     & * \\
    
    \midrule
    U3 &       &       &       &       &       &       &  \\
    Acceptance: Email address &       &       &       &       &       &       &  &  \\
    
    \hspace{1em}Online banking & 1.1443 & 0.5643 &       & 0.8556 & *     &       &       & * \\
    \hspace{1em}Online shop & 0.6945 & 0.7066 &       & *     & *     &       &       & * \\
    \hspace{1em}Email service & 2.0574 & 0.3575 &       & 0.4573 & *     &       &       & * \\
    \hspace{1em}Social network & 6.1693 & \textbf{0.0457} &       & 0.1945 & 0.0580 &       &       & * \\
    \hspace{1em}Online storage & 1.8262 & 0.4013 &       & *     & 0.5307 &       &       & * \\
    \hspace{1em}Video website & 3.0306 & 0.2197 &       & *     & 0.4030 &       &       & 0.3567 \\
    \hspace{1em}News website & 11.3867 & \textbf{0.0034} &       & *     & \textbf{0.0029} &       &       & \textbf{0.0491} \\
    
    \midrule
    Acceptance: Phone number &       &       &       &       &       &       &  &  \\
    \hspace{1em}Online banking & 3.1975 & 0.2021 &       & *     & 0.3644 &       &       & 0.3644 \\
    \hspace{1em}Online shop & 1.6267 & 0.4434 &       & 0.6720 & *     &       &       & * \\
    \hspace{1em}Email service & 2.9798 & 0.2254 &       & 0.2992 & 0.6091 &       &       & * \\
    \hspace{1em}Social network & 0.6346 & 0.7281 &       & *     & *     &       &       & * \\
    \hspace{1em}Online storage & 4.3858 & 0.1116 &       & 0.1495 & *     &       &       & 0.3182 \\
    \hspace{1em}Video website & 0.9300 & 0.6281 &       & *     & *     &       &       & * \\
    \hspace{1em}News website & 0.4674 & 0.7916 &       & *     & *     &       &       & * \\
    \midrule
    Acceptance: App &       &       &       &       &       &       &  & * \\
    \hspace{1em}Online banking & 1.3492 & 0.5094 &       & *     & 0.7563 &       &       & * \\
	\hspace{1em}Online shop & 0.3868 & 0.8241 &       & *     & *     &       &       & * \\
	\hspace{1em}Email service & 0.6302 & 0.7297 &       & *     & *     &       &       & * \\
	\hspace{1em}Social network & 0.5397 & 0.7635 &       & *     & *     &       &       & * \\
	\hspace{1em}Online storage & 0.1353 & 0.9346 &       & *     & *     &       &       & * \\
	\hspace{1em}Video website & 0.0880 & 0.9570 &       & *     & *     &       &       & * \\
	\hspace{1em}News website & 2.1923 & 0.3342 &       & 0.5509 & 0.6573 &       &       & * \\
    \midrule
    S1 &       &       &       &       &       &       &  \\
    \hspace{1em}How secure in general & 22.2597 & \textbf{\textless 0.0001} & \textbf{0.0002} & *     & *     & \textbf{0.0013} & \textbf{0.0017} & * \\
    \hspace{1em}How secure compared to PW & 7.0254 & \textbf{0.0298} &       & 0.1591 & \textbf{0.0336} &       &       & * \\
    \midrule
    S2: Protection: General & 20.3219 & \textbf{\textless 0.0001} & \textbf{\textless 0.0001} & *     & *     & \textbf{0.0126} & \textbf{0.0113} & * \\
    \midrule
    S3: Protection: Scenarios &       &       &       &       &       &       &  \\
    \hspace{1em}Online banking & 7.1264 & {0.0680} & \textbf{0.0499} & *     & *     & 0.6822 & 0.7232 & * \\
    \hspace{1em}Online shop & 14.0265 & \textbf{0.0029} & \textbf{0.0057} & *     & *     & \textbf{0.0105} & 0.1386 & * \\
    \hspace{1em}Email service & 5.2338 & 0.1555 & 0.3046 & *     & *     & 0.3674 & 0.5903 & * \\
    \hspace{1em}Social network & 14.2490 & \textbf{0.0026} & *     & {0.0644} & *     & \textbf{0.0014} & 0.4721 & 0.3284 \\
    \hspace{1em}Online storage & 7.5093 & {0.0573} & 0.2652 & *     & *     & 0.1483 & 0.1013 & * \\
    \hspace{1em}Video website & 5.4783 & 0.1399 & *     & 0.3389 & *     & 0.2245 & *     & * \\
    \hspace{1em}News website & 0.0853 & 0.9935 & *     & *     & *     & *     & *     & * \\
    
    \end{tabular}%
	}
  \label{tab:omnibusposthoc}%
\end{table}%

 \end{document}